\documentclass[twoside,twocolumn,9pt]{article}
\usepackage{extsizes}
\usepackage[super,sort&compress,comma]{natbib} 
\usepackage[version=3]{mhchem}
\usepackage[left=1.5cm, right=1.5cm, top=1.785cm, bottom=2.0cm]{geometry}
\usepackage{balance}
\usepackage{times,mathptmx}
\usepackage{sectsty}
\usepackage{graphicx} 
\usepackage{lastpage}
\usepackage[format=plain,justification=justified,singlelinecheck=false,font={stretch=1.125,small,sf},labelfont=bf,labelsep=space]{caption}
\usepackage{float}
\usepackage{fancyhdr}
\usepackage{fnpos}
\usepackage[english]{babel}
\usepackage{array}
\usepackage{droidsans}
\usepackage{charter}
\usepackage[T1]{fontenc}
\usepackage[usenames,dvipsnames]{xcolor}
\usepackage{setspace}
\usepackage[compact]{titlesec}
\usepackage{multirow,diagbox,tabularx,blindtext}
\usepackage{mathtools,amssymb,lipsum}
\usepackage{cuted}

\usepackage{epstopdf}
\DeclareSymbolFont{matha}{OML}{txmi}{m}{it}
\DeclareMathSymbol{\varv}{\mathord}{matha}{118}

\definecolor{cream}{RGB}{222,217,201}

\begin{document}

\pagestyle{fancy}
\thispagestyle{plain}
\fancypagestyle{plain}{

\fancyhead[C]{\includegraphics[width=18.5cm]{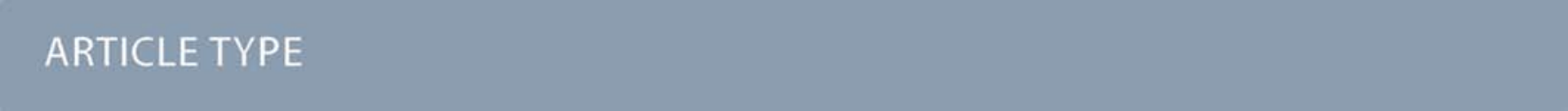}}
\fancyhead[L]{\hspace{0cm}\vspace{1.5cm}\includegraphics[height=30pt]{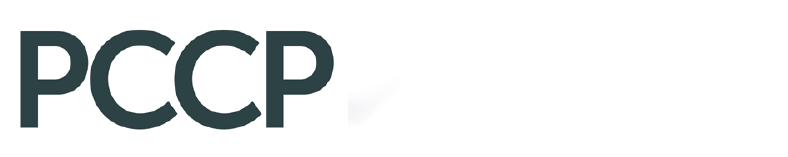}}
\fancyhead[R]{\hspace{0cm}\vspace{1.7cm}\includegraphics[height=55pt]{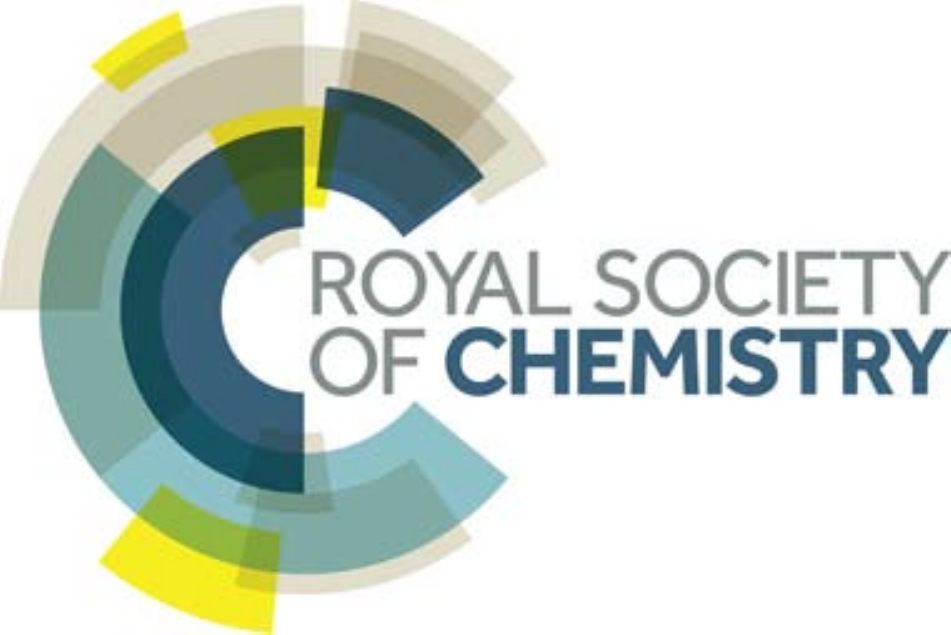}}
\renewcommand{\headrulewidth}{0pt}
}

\makeFNbottom
\makeatletter
\renewcommand\LARGE{\@setfontsize\LARGE{15pt}{17}}
\renewcommand\Large{\@setfontsize\Large{12pt}{14}}
\renewcommand\large{\@setfontsize\large{10pt}{12}}
\renewcommand\footnotesize{\@setfontsize\footnotesize{7pt}{10}}
\makeatother

\renewcommand{\thefootnote}{\fnsymbol{footnote}}
\renewcommand\footnoterule{\vspace*{1pt}%
\color{cream}\hrule width 3.5in height 0.4pt \color{black}\vspace*{5pt}} 
\setcounter{secnumdepth}{5}

\makeatletter 
\renewcommand\@biblabel[1]{#1}            
\renewcommand\@makefntext[1]%
{\noindent\makebox[0pt][r]{\@thefnmark\,}#1}
\makeatother 
\renewcommand{\figurename}{\small{Fig.}~}
\sectionfont{\sffamily\Large}
\subsectionfont{\normalsize}
\subsubsectionfont{\bf}
\setstretch{1.125} 
\setlength{\skip\footins}{0.8cm}
\setlength{\footnotesep}{0.25cm}
\setlength{\jot}{10pt}
\titlespacing*{\section}{0pt}{4pt}{4pt}
\titlespacing*{\subsection}{0pt}{15pt}{1pt}

\fancyfoot{}
\fancyfoot[LO,RE]{\vspace{-7.1pt}\includegraphics[height=9pt]{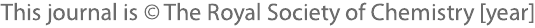}}
\fancyfoot[CO]{\vspace{-7.1pt}\hspace{11.9cm}\includegraphics{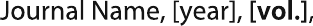}}
\fancyfoot[CE]{\vspace{-7.2pt}\hspace{-13.2cm}\includegraphics{head_foot/RF}}
\fancyfoot[RO]{\footnotesize{\sffamily{1--\pageref{LastPage} ~\textbar  \hspace{2pt}\thepage}}}
\fancyfoot[LE]{\footnotesize{\sffamily{\thepage~\textbar\hspace{4.65cm} 1--\pageref{LastPage}}}}
\fancyhead{}
\renewcommand{\headrulewidth}{0pt} 
\renewcommand{\footrulewidth}{0pt}
\setlength{\arrayrulewidth}{1pt}
\setlength{\columnsep}{6.5mm}
\setlength\bibsep{1pt}

\makeatletter 
\newlength{\figrulesep} 
\setlength{\figrulesep}{0.5\textfloatsep} 

\newcommand{\topfigrule}{\vspace*{-1pt}%
\noindent{\color{cream}\rule[-\figrulesep]{\columnwidth}{1.5pt}} }

\newcommand{\botfigrule}{\vspace*{-2pt}%
\noindent{\color{cream}\rule[\figrulesep]{\columnwidth}{1.5pt}} }

\newcommand{\dblfigrule}{\vspace*{-1pt}%
\noindent{\color{cream}\rule[-\figrulesep]{\textwidth}{1.5pt}} }

\makeatother

\twocolumn[
  \begin{@twocolumnfalse}
\vspace{3cm}
\sffamily
\begin{tabular}{m{4.5cm} p{13.5cm} }

\includegraphics{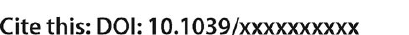} & \noindent\LARGE{\textbf{Generalization of the Elastic Network model for the study of large conformational changes in biomolecules$^\dag$}} \\
\vspace{0.3cm} & \vspace{0.3cm} \\

 & \noindent\large{Adolfo B. Poma,$^{\ast}$\textit{$^{a}$} Mai Suan Li,\textit{$^{a}$} and Panagiotis E. Theodorakis$^{\ast}$\textit{$^{a}$}} \\

\includegraphics{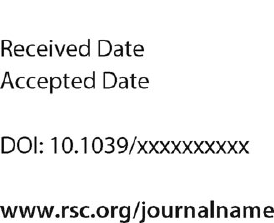} & \noindent\normalsize{The Elastic Network (EN) is a prime model that describes the long-time dynamics of
biomolecules. 
However, the use of harmonic potentials renders this model 
insufficient for studying large conformational changes of proteins (\textit{e.g.} stretching of proteins, folding and thermal unfolding). Here, we extend the capabilities
of the EN model by using a harmonic approximation described by 
Lennard--Jones (LJ) interactions for far contacts and
native contacts obtained from the standard overlap criterion as in the case of G\=o-like models.
While our model is validated against the EN model by reproducing the equilibrium properties for a number of proteins,
	we also show  that the model is suitable for the study of large conformation changes by providing various examples.
In particular, this is illustrated on the basis of pulling simulations that predict with high accuracy the experimental data on the rupture force of the studied proteins. Furthermore, in the case of DDFLN4 protein, our pulling simulations highlight
	the advantages of our model with respect to G\=o-like approaches, where the latter fail
	to reproduce previous results obtained by all-atom
simulations that predict an additional characteristic peak for this protein. In addition, folding simulations of small peptides yield different folding times for $\alpha$-helix and $\beta$-hairpin, in agreement with experiment, in this way providing further opportunities for the application of our model in studying large conformational changes of proteins. In contrast to the EN
model, our model is suitable for both normal mode analysis and molecular dynamics simulation.
We anticipate that the proposed model will find applications in a broad range of problems in biology, including, among others, protein folding and thermal unfolding.} \\

\end{tabular}

 \end{@twocolumnfalse} \vspace{0.6cm}

  ]

\renewcommand*\rmdefault{bch}\normalfont\upshape
\rmfamily
\section*{}
\vspace{-1cm}


\footnotetext{\textit{$^{a}$~Institute of Physics, Polish Academy of Sciences, Al. Lotnik\'ow 32/46, 02-668 Warsaw, Poland; E-mail: poma@ifpan.edu.pl, panos@ifpan.edu.pl}}

\footnotetext{\dag~Electronic Supplementary Information (ESI) available: Harmonic approximation for the Lennard--Jones potential. Pulling results for titin and sequence of transition states during pulling simulation. See DOI: 10.1039/cXCP00000x/}



\section{Introduction}\label{sec:intro}
One of the main goals in the computer simulation arena of biomolecules is to build
the simplest yet computationally most efficient models able to reproduce accurately and predict faithfully dynamic and structural properties of proteins. 
A most prime example of this is the elastic network (EN) \cite{Tirion1996},  which reproduces
well the low-frequency motion (long-time dynamics) of proteins. The EN has been also employed for modelling other important biomolecules such as DNA \cite{Setny2013}, RNA \cite{Zimmermann2014,Pinamonti2015}, graphene sheet \cite{KiM2014}, and cellulose fibers \cite{Glass2012}, providing
information on their equilibrium dynamics, the influence of the native-structure topology
on their stability, the localization properties of protein fluctuations
or the definition of protein domains \cite{Cui2006}. 
Although a number of similar models have subsequently appeared in the literature
and various improvements have been suggested \cite{Bahar1997,Bahar_PRL,Hinsen1998,Hinsen1999,Atilgan2001,Tama2001},
the EN still remains the standard model having attracted particular interest due to its simplicity and ability to provide realistic frequency data \cite{Cui2006}. The use of EN model for studying processes that involve large conformational changes of proteins is a current challenge though, in practice due to the required numerical complexity. Therefore, several methodologies have been developed to tackle this problem. For example, certain approaches are based on the update of the connectivity or Kirchoff matrix during a linear interpolation between two known protein states \cite{kiM2002efficient,feng2009energy,das2014exploring,tekpinar2010predicting}. However, this approximation fails when the two states are unknown or when one or both of these states are represented
by an ensemble of equivalent configurations (\textit{e.g.} unfolded state).

\begin{figure}[t]
 \centering
   \includegraphics[scale =0.15] {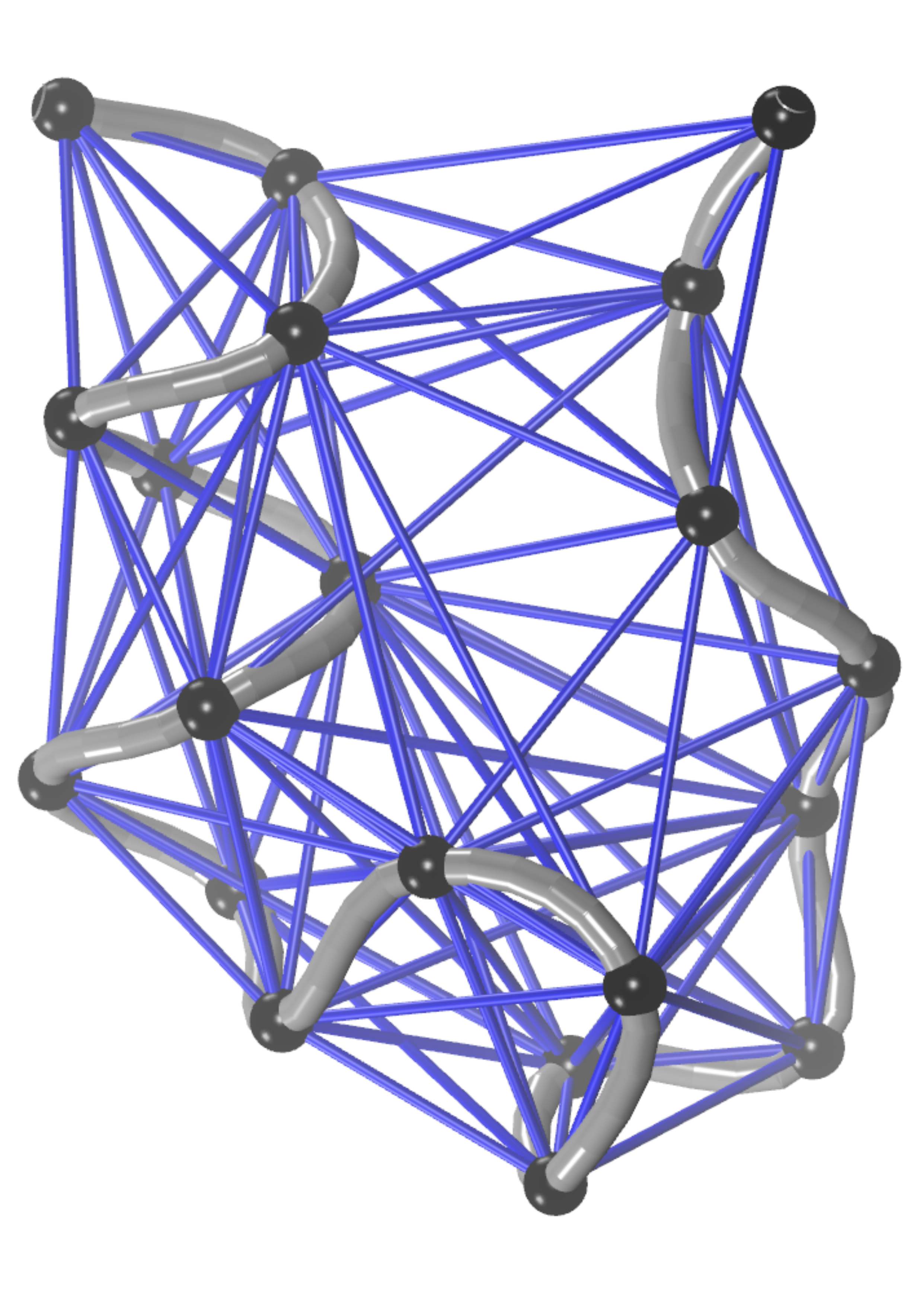}\hspace{1cm}
   \includegraphics[scale =0.15]{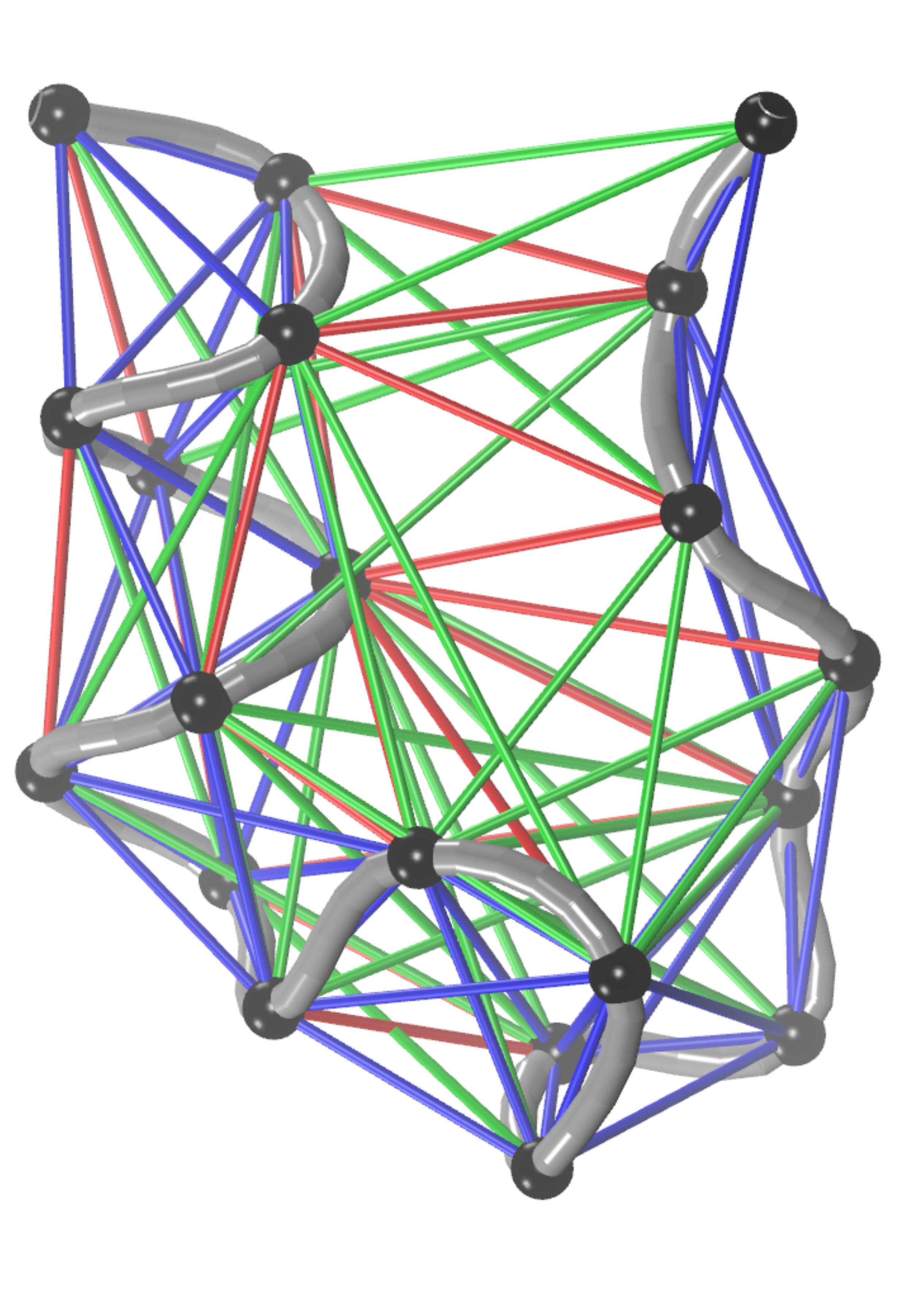}
   \caption{\small{Left panel shows the EN model for a tryptophan-cage motif (PDB ID: 1L2Y), the ``unbreakable'' harmonic EN contacts are shown in blue color. Right panel shows the GEN model, where a subset of the EN contacts are formulated as G\=o contacts (in red) and far contacts by an effective harmonic term based on the LJ potential (in green). The EN and GEN models do not assume \textit{a priori} C$_{\rm \alpha}$-backbone connectivity. Hence, the
   			tube representation of the C$_{\rm \alpha}$ backbone in grey serves only as a guide for the eye. 
   		Still, consecutive C$_{\rm \alpha}$ atoms are connected by EN harmonic bonds, because they are 
   		within the cutoff distance of the EN model (see main text for further details).}}
   \label{intro}
 \end{figure}

The EN model is based only on a single-parameter harmonic potential between residues that are represented in the model
by the C$_{\rm \alpha}$ atoms. In this model, the harmonic interaction is introduced when two residues overlap, \textit{i.e.} the van der Waals radii augmented by a cutoff distance of any pairs of atoms belonging to different residues overlap (see Table~\ref{vdWradii}). Here, the harmonic approximation of EN contacts models the interaction between C$_{\rm \alpha}$ atoms in the native state, such as the electrostatic and van der Waals interactions, as well as the covalent bonds along the backbone of C$_{\rm \alpha}$ atoms  (Fig.~\ref{intro}). On the one hand, the harmonic approximation is incompatible with the dissociation of native contacts for certain processes involving large conformational changes in biomolecules.
On the other hand, some important advantages of the EN from the modeling point of view are avoiding: the use of computationally
expensive simulations based on all-atom force-fields and the necessity of including 
\textit{ad hoc} backbone stiffness in the model. The latter is generally implemented in coarse-grained models based on C$_{\rm \alpha}$ atoms \cite{Clementi2000,Karanicolas2002,Poma2017} to mimic the all-atom description of proteins, which is based on harmonic interactions used to maintain bonds and bond and dihedral angles. 
Technically, the EN model \cite{Tirion1996} is suitable for Normal Mode Analysis (NMA), which requires the calculation of the second-derivative (or Hessian) matrix. However, this step requires substantial computer memory and processing power to perform the matrix diagonalization, which becomes a severe bottleneck in the study of very large complexes. Moreover, one typically ignores molecular-interaction details \cite{van2004normal} at this coarse-grain level of description, which is an approach that has been shown to work better in cases where the protein is packed uniformly \cite{bagci2003origin}. While an initial energy-minimization step is needed to satisfy the harmonic pairs (\textit{e.g.} steepest descent \cite{fletcher1963rapidly} and conjugate gradient \cite{kershaw1978incomplete} methods), this is not necessary for small systems \cite{periole2009combining}.\\
Despite the aforementioned advantages of the EN, this model cannot be currently used for studying certain processes that involve large conformational changes of proteins
apart from a few previous attempts that require \textit{a priori} knowledge of the protein states \cite{kiM2002efficient,feng2009energy,das2014exploring,tekpinar2010predicting},
due to the presence of the harmonic bonds. In the following, we overcome this barrier and enhance
the number of possible applications of the EN, for example, protein stretching \cite{Rief1997,Kellermayer1997,Sulkowska2008,Poma2017,Kumar2010}, prediction of elastic properties \cite{Becker2003} without the assumption of continuum theory \cite{Landau1986}, characterization of folding pathways \cite{Jackson1998} denaturation due to high temperature \cite{Benjwal2006}, pressure \cite{pressure} and surfactants \cite{surfactant}, as well as
denaturation processes at interfaces (\textit{e.g.} air--water and oil--water interfaces) based on simple CG potentials \cite{yani}. Here, we propose a model where 
harmonic interactions beyond the nearest neighbors (distance in sequence larger than three)
are substituted by harmonic effective terms approximated by the Lennard--Jones (LJ) potential \cite{Poma2015}. In addition,
our model assumes LJ contacts obtained from a contact map based on the standard overlap 
criterion (Fig.~\ref{intro}) \cite{Cieplak2003,Tsai1999}, namely, we determine native contacts based on the overlap (OV) of vdW radii of heavy atoms (N, C, C$_{\rm \alpha}$ and O). For 
the determination of the contact map one considers each residue as a cluster of spheres. In this case, the radii of the spheres are equal to the van der Waals radii enhanced by a 
factor of about 25\% (see Table \ref{vdWradii}). In addition, our model does not require bending- 
and dihedral-angle parameters along the C$\rm \alpha$ backbone in the native state.
Our model is based only on a single parameter as the EN model does, but it enables the 
study of large-conformational changes in biomolecules due to the use of the contact map.
In this respect, we anticipate that our work opens the door for the use of EN in a wider range of applications, for example, folding, thermal unfolding,
denaturation of proteins at interfaces, \textit{etc}. 

In the following sections, we provide details about our model and methodologies. 
After validating the proposed model with the EN model for a number of proteins, we present various
studies that involve large conformational changes of proteins highlighting model's advantages in comparison with the EN and G\=o-like models.

\begin{table*}[t]
\centering
\caption{\small{List of vdW radii for heavy atoms used to determine the presence
of the native contacts in proteins, sugars, and the sugar--protein complex.
The third column refers to proteins. The values are taken from ref. \cite{Tsai1999}.
The radii of the enlarged spheres are defined as
the vdW radii multiplied by 1.24 to account for attraction \cite{64models}.}}
\minipage{1.0\textwidth}
\resizebox{0.75\linewidth}{!}{%
\hspace*{4.0cm}
\begin{tabular}{|c|c|c|c|}
\hline
\hspace*{0.3cm}{\large No.}\hspace*{0.3cm}&\hspace*{0.3cm}{\large Atomic Group}\hspace*{0.3cm}& \hspace*{0.3cm}{\large vdW radius [{\AA}]}\hspace*{0.3cm} & \hspace*{0.3cm}{\large enlarged radius [{\AA}]}\hspace*{0.3cm} \\
\hline
{\large 1}&{\large C$_3$H$_0$} & {\large 1.61} & {\large 2.00} \\
\hline
{\large 2}&{\large C$_3$H$_1$} & {\large 1.76} & {\large 2.18}\\
\hline
{\large 3}&{\large C$_4$H$_1$} & {\large 1.88} & {\large 2.33} \\
\hline
{\large 4}&{\large C$_4$H$_2$} & {\large 1.88} & {\large 2.33} \\
\hline
{\large 5}&{\large C$_4$H$_3$} & {\large 1.88} & {\large 2.33} \\
\hline
{\large 6}&{\large N$_3$H$_0$} & {\large 1.64} & {\large 2.03}\\
\hline
{\large 7}&{\large N$_3$H$_1$} & {\large 1.64} & {\large 2.03} \\
\hline
{\large 8}&{\large N$_3$H$_2$} & {\large 1.64} & {\large 2.03} \\
\hline
{\large 9}&{\large N$_4$H$_3$} & {\large 1.64} & {\large 2.03}\\
\hline
{\large 10}&{\large O$_1$H$_0$} & {\large 1.42} & {\large 1.76}  \\
\hline
{\large 11}&{\large O$_2$H$_1$} & {\large 1.46} & {\large 1.81} \\
\hline
{\large 12}&{\large S$_2$H$_0$} & {\large 1.77} & {\large 2.19} \\
\hline
{\large 13}&{\large S$_2$H$_1$} & {\large 1.77} & {\large 2.19} \\
\hline
\end{tabular}
}
\endminipage\hfill%

\label{vdWradii}
\end{table*}

\section{\label{sec:methods} Materials and Methods}
\subsection{Generalized elastic network model }
In the case of EN, if any two heavy atoms belonging to different residues are within
a distance $r_{\rm c} = \varv dW_{\rm i}+ \varv dW_{\rm j} + R_c$, then the two residues form an EN contact
and a harmonic potential that connects the C$_{\rm \alpha}$ positions of these residues is applied. 
$R_{\rm c}$ is simply a cutoff distance and $\varv dW_{\rm i}$ and $\varv dW_{\rm j}$ are the van der Waals
radii of heavy atoms belonging to residues $\rm i$ and $\rm j$ (see Table~\ref{vdWradii}). 
In this case, the harmonic potential has the form $V_{\rm h}=C(r_{\rm ij}-r_{\rm ij}^{0})^{2}$, where $r_{\rm ij}^{0}$ is the
distance between C$_{\rm \alpha}$ atoms in the native structure of the protein and $C$ is a constant indicating the strength of the harmonic potential. The energy scale associated to the EN model\cite{Tirion1996} is given by $\epsilon_{\rm EN}=CR_{\rm c}^{2}$.
In the case of our model, we use the same guideline for defining contacts due to EN. They contribute both to bonded and nonbonded energy of the Hamiltonian. The energy can be written as,
\begin{equation} 
 \mathcal{H}=\sum_{|i-j| < 4}C(r_{\rm ij}-r_{\rm ij}^{0})^{2}+\sum_{|i-j|>3}\mathcal{U}_{\rm ij}^{\text{nb}}
\end{equation}
The first term is the standard harmonic potential associated with the EN model while the second term is the nonbonded contribution,
which enables large conformational changes in the protein. This second term is defined as follows:
\vspace{1in}
\begin{strip}
\begin{equation} 
 \mathcal{U}_{\rm ij}^{\text{nb}}=\begin{cases}
               U_{\rm LJ}(\epsilon_{\rm cm})&\text{if a native contact forms according to the OV criterion as in G\=o-like models}\\
               U_{\rm LJ}(\epsilon_{\rm harm})& \text{If a non-native contact forms according to the cutoff distance as in the EN model }
            \end{cases}
\end{equation}
\end{strip}

The native contacts found by the OV criterion are described by a G\=o-like model \cite{go1978respective,go1981noninteracting} with LJ interactions ($U_{\rm LJ}(\epsilon_{\rm cm})$).  In addition, we use an effective-harmonic term based on the LJ potential [$U_{\rm LJ}(\epsilon_{\rm harm})$] for contacts between residues with a distance in sequence
larger than three. The LJ potential reads
\begin{equation}\label{eq:LJpotential}
U_{\rm LJ}(\epsilon_{\rm ij}) = 4\epsilon_{\rm ij} \left[  \left(\frac{\sigma_{\rm ij}}{r_{\rm ij}}
\right)^{12} - \left(\frac{\sigma_{\rm ij}}{r_{\rm ij}}  \right)^{6}    \right],
\end{equation}
where $r_{\rm ij}$ is the distance between any pair of $\rm i$ and $\rm j$ C$_{\rm \alpha}$ atoms
in the system.
The relation between the effective harmonic term and the strength of the 
LJ potential $\epsilon_{\rm ij}$ is simply described by the formula\cite{Poma2015}: $\epsilon_{\rm ij}=\epsilon_{\rm harm}$, where $\epsilon_{\rm harm}=C\sigma_{\rm ij}^{2}36^{-1}(2^{2/3})$ (see SI),
$\sigma_{\rm ij}= 2^{-1/6}r_{\rm ij}^{0}$. Hence, one infers about the $\epsilon_{\rm harm}$ from  $C$ and $r_{\rm ij}^{0}$. 
In addition, we include contacts by using a contact map based on the standard overlap criterion \cite{Cieplak2003,Tsai1999}. The
latter contacts are represented by LJ potentials $U_{\rm LJ}(\epsilon_{\rm cm})$ \cite{Hoang2000,Sulkowska2007}. In this case, the 
strength of interaction is independent of the distance and equal to $\epsilon_{\rm ij}=\epsilon_{\rm cm}$ for any pair of residues in contact, where $\epsilon_{\rm cm}$ is the unit of energy, and $\sigma_{\rm ij}=2^{-1/6}r_{\rm ij}^{0}$. Here, the subindex ``cm'' denotes the ``contact map" obtained by the overlap criterion. Moreover, the latter contacts apply only for residues at a sequential distance larger than three.
If a native contact coincides with a harmonic effective contact, then the native contact [$U_{\rm LJ}(\epsilon_{\rm cm})$] remains and the harmonic contact [$U_{\rm LJ}(\epsilon_{\rm harm})$] is
removed. 
An important characteristic of the model is the proper balance of the energy scales assigned between contacts for the C$_{\rm \alpha}$ backbone and those that represent the non-bonded contributions. In particular, there are three energy scales: 
the EN-type interactions ($\epsilon_{\rm EN}$), the native ($\epsilon_{\rm cm}$) 
and the effective harmonic-type interactions ($\epsilon_{\rm harm}$). 
$\epsilon_{\rm EN}=CR_{\rm c}^{2}$, which corresponds to 12.250 kJmol$^{-1}$ 
for $R_{\rm c}=0.35$ nm, and $\epsilon_{\rm cm}= 6.276~$kJmol$^{-1}$, which reflects 
the strength of hydrogen bond in proteins \cite{Poma2015}. 
$\epsilon_{\rm harm}$ is about 2.0 kJmol$^{-1}$.
Henceforth, we will simply refer to our model as Generalized Elastic Network (GEN) model.
We have also investigated other possibilities, such as
excluding the effective harmonic interactions $U_{\rm LJ}(\epsilon_{\rm harm})$  and substituting
all effective harmonic terms with standard LJ potentials
($\epsilon_{\rm ij}=\epsilon_{\rm cm}$, ``M1'' model) or eliminating all   contacts beyond 1--4 [$U_{\rm LJ}(\epsilon_{\rm harm})$]  and keeping
only the native contacts [$U_{\rm LJ}(\epsilon_{\rm cm})$] based on the overlap criterion (``M2'' model) . Finally, we have also considered
the case that we have all EN harmonic bonds irrespective of the sequential distance between residues as in the standard
EN model, but those contacts that coincide with the native 
contacts derived from the overlap criterion will be substituted by LJ potentials $U_{\rm LJ}(\epsilon_{\rm cm})$ resulting
in what we will simply refer to as the ``M3'' model. In the following, we will discuss these models
for the sake of comparison with the GEN and EN models on the basis of the different number of harmonic, effective harmonic and G\=o contacts (see Table~\ref{table1}).

\subsection{Normal mode analysis and pulling simulations}

We used the GROMACS package \cite{Berendsen1995,vanderSpoel2005,Hess2008,Pronk2013} to perform
standard Normal Mode Analysis (NMA) \cite{Cui2006}. The output of NMA is independent
(normal) modes (harmonic motions) characterized by an eigenvalue 
(characteristic frequency). Each normal mode acts as a simple harmonic oscillator of
a concerted motion of atoms without moving the center of mass with all atoms passing
through their equilibrium position at the same time. Moreover, normal modes 
resonate independently and can be obtained directly by data obtained from 
vibrational spectroscopy. In practice, the normal modes are the eigenvectors of the
Hessian matrix, which represents the force constants between every possible pair of residues in the system in all directions of the Cartesian coordinate system.
 
The mean square fluctuations of the C$_{\rm \alpha}$ atoms are calculated from the normal modes as follows:
\begin{equation}
\langle\Delta r_{\rm j}^2\rangle=k_{\rm B}T\sum_{\rm i}\frac{|\vec{a}_{\rm ij}|^2}{\omega_{\rm i}^2},
\end{equation}
here, $\vec{a}_{\rm ij}$ is the vector of the projections of the $\rm i$-th eigenvector of the normal modes set with frequency $\omega_i$ on the Cartesian components of the displacement vector for the $\rm j$-th C$_{\rm \alpha}$ atom, $k_{\rm B}$ is the Boltzmann constant, and, $T$, the reference temperature. The B-factor related to the expected
residue fluctuations is calculated by the following relation
\begin{equation}
B_{\rm j} = \frac{8\pi^2}{3} \langle\Delta r_{\rm j}^2\rangle.
\end{equation}
To perform the protein stretching, we used Molecular Dynamics (MD) simulation in the NVT 
ensemble. The time step was 0.01 ps and the protein was pulled along the end-to-end vector connecting the C$_{\rm \alpha}$-atoms from the N- and C-termini and the reaction coordinate is the displacement of the pulling spring. Moreover, additional beads have been attached to those C$_{\rm \alpha}$-atoms with the spring constant being 37.6 kJ mol$^{-1}$ nm$^{-2}$, which is a typical value of the  Atomic Force Microscopy (AFM) cantilever stiffness in protein stretching studies \cite{Carrion-Vazquez1999}. Each system was pulled over the course of 10$^{7}$ ps with a velocity of 10$^{-2}$ m/s. Although this value is still far from the experimental value of cantilever velocity\cite{marszalek1999mechanical} ($\sim 10^{-6}$ m/s ), it gives comparable results with experiments  showing the intrinsic speedup associated to the smoothing of the potential energy landscape, which is typical for coarse-grain methods.

\section{\label{sec:results} Results and Discussion}

In this section, we first validate the GEN model in terms of B-factors by comparing with data obtained from the EN model. Then, using this model we have performed stretching and folding simulations, in this way providing two illustrative examples of
 large conformation changes in proteins. 
  
\subsection{Validation of the GEN model}

\begin{figure}[ht!]
 \centering
   \includegraphics[scale =0.45]{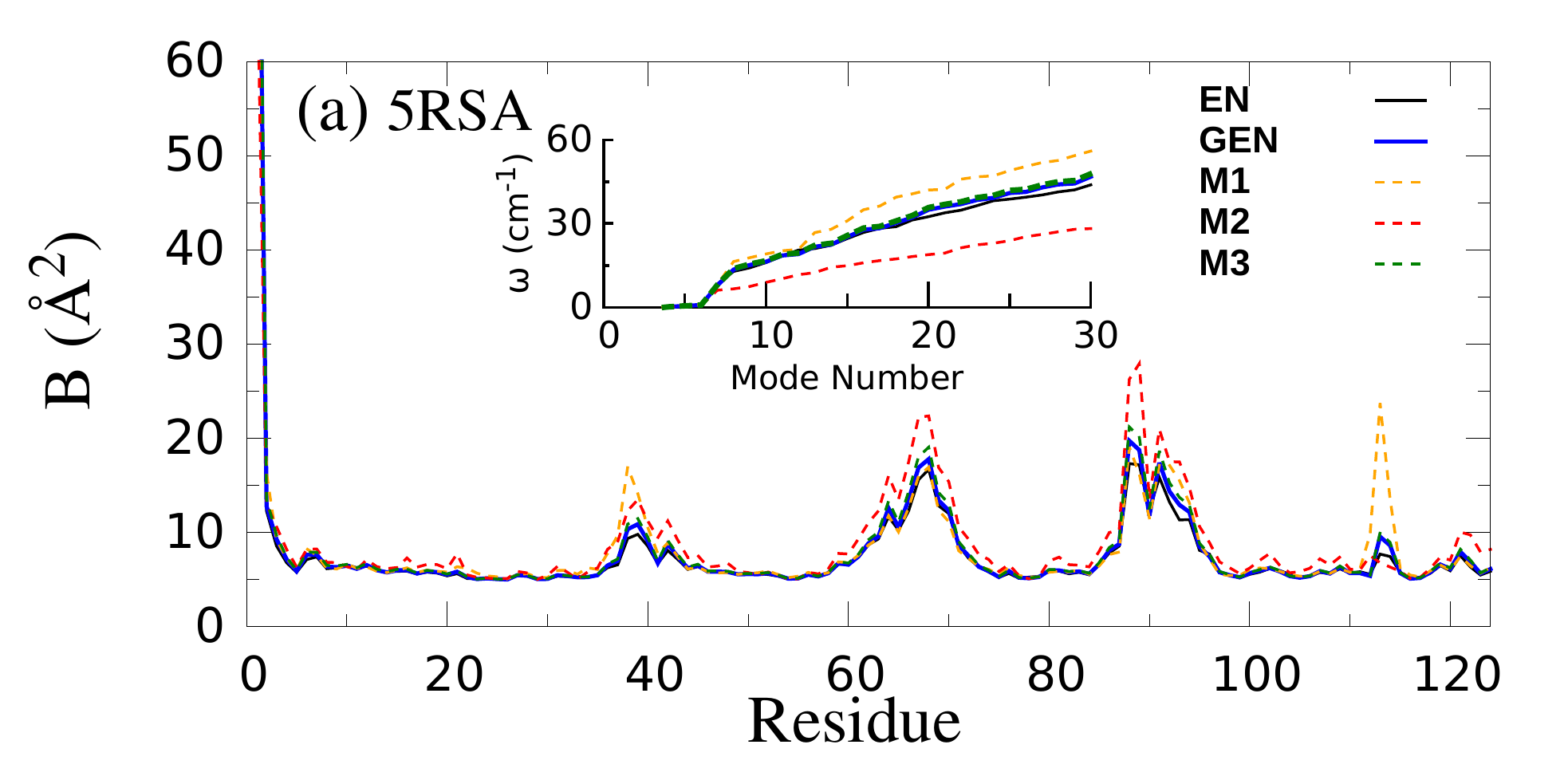}
   \includegraphics[scale =0.45]{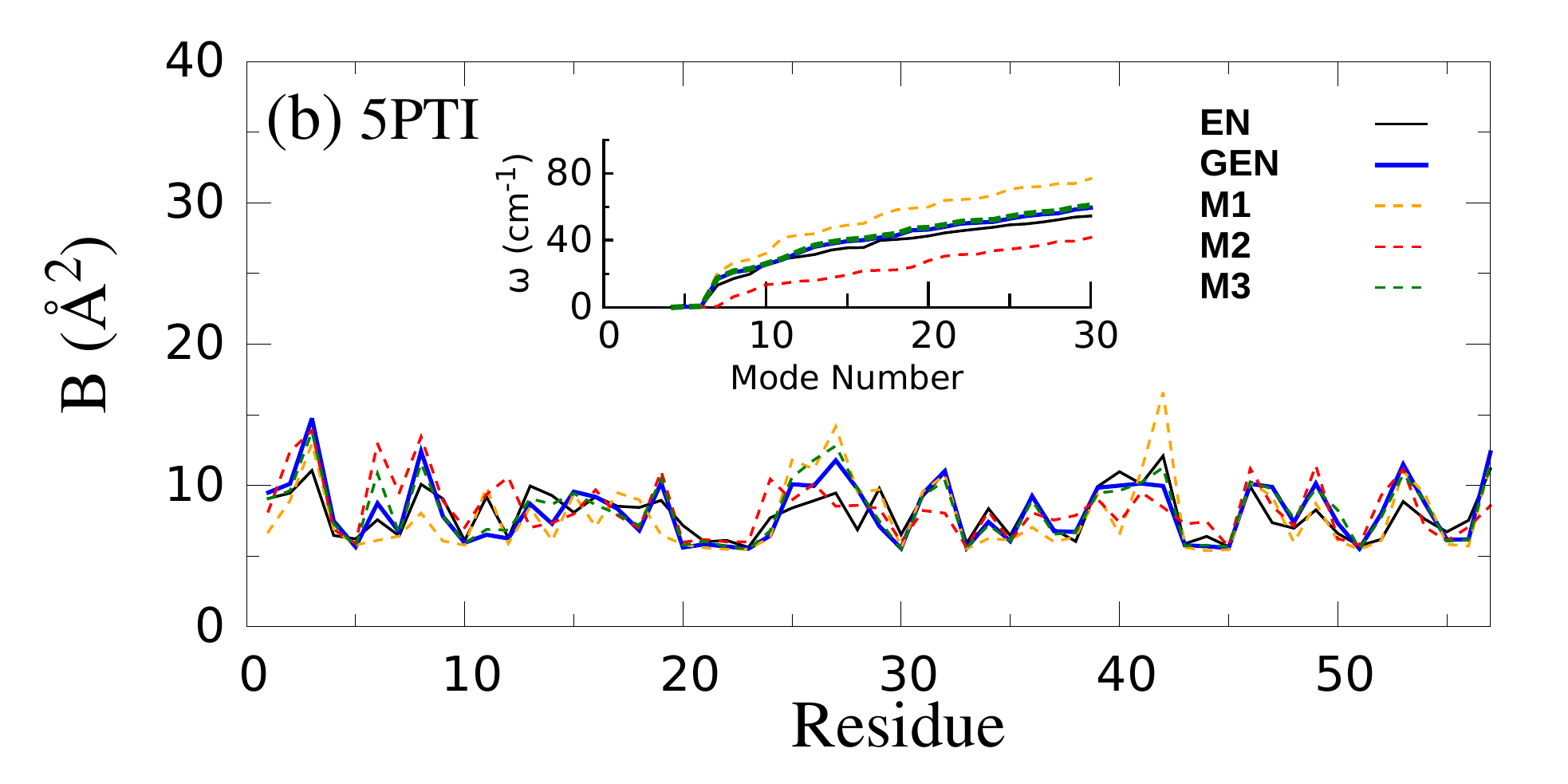}
   \includegraphics[scale =0.45]{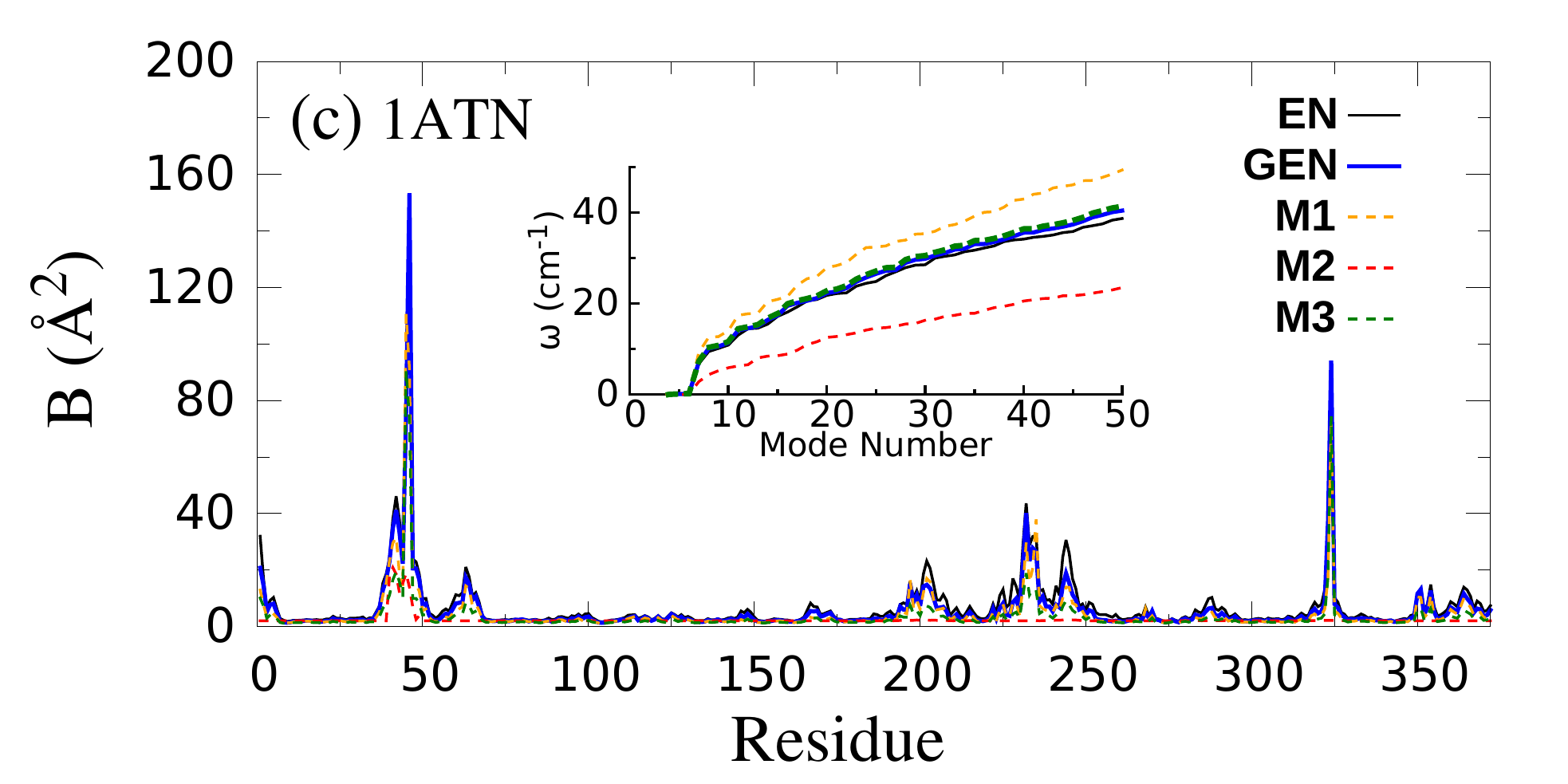}
   \caption{\small{The B-factors for the lowest 30 modes for the proteins with PDB ID: 5RSA (Ribonuclease-A, a) , 5PTI (Bovine Pacreatic Trypsin Inhibitor, b), and 1ATN (G-actine, c).
   The results obtained from different models are illustrated as indicated. Here, $R_{\rm c}=$0.35 nm. Insets show the frequency $\omega$ (cm$^{-1}$) for the lowest 30 modes.}}
   \label{betafactor_5RSA_5PTI}
 \end{figure}

To validate our model, we have calculated the B-factors for several target proteins by using NMA \cite{Bahar2005}, which are proportional to the mean square fluctuations of atom positions. We juxtaposed our results with those of Tirion \cite{Tirion1996}, which were also obtained by using the same EN approach (Fig.~\ref{betafactor_5RSA_5PTI}). The results shown in Fig.~\ref{betafactor_5RSA_5PTI} were
obtained for a cutoff distance of $R_{\rm c}=0.35$ nm, but, for other models considered in this study, we have also investigated different cutoffs, namely, $R_{\rm c}=$ 0.2 - 0.85 nm. In our case, the best correlation was obtained for $R_{\rm c}=$0.35 nm. Our data manifests an excellent agreement with the EN model showing that the GEN model reproduces closely the properties of the EN, which we have also confirmed for all proteins discussed in ref.~\cite{Tirion1996} providing a theoretical validation of the GEN model in the case of the chosen set of proteins. Another  EN model, the so-called Gaussian Network model (GNM)\cite{Bahar1997,Bahar_PRL}, is able to reproduce closer the experimental results of B-factor related to larger protein complexes. 
 However, this may not be the case for atomic fluctuations derived from all-atom simulation \cite{rmsfAA}. Moreover, the GNM cannot be used in simulation and, therefore, the one-to-one correspondence with short-time atomic fluctuations is not conceived in this formalism. In addition, this model introduces additional concepts from the elastic theory of random polymer network \cite{Flory1976} and it is found to be more appropriate to reproduce available experimental fluctuation data reported by the B-factors in the case of G-actine. 
Yet, even models based on this assumption are not reliable for describing large conformational changes as they rely by construction on the ``unbreakable'' harmonic bonds.

\begin{table*}[t]
\centering
\caption{\small{\label{table1}Total number of bonds/contacts for different models as indicated. Here, the columns are as follows: ``EN'' indicates the 
    number of EN bonds, ``cm'' the number of native contacts by using the contact map (overlap criterion), and the ``Eff. Harm. (LJ)'' indicates
    the number of effective-harmonic contacts. $R_{\rm c}=$0.35 nm.}}
\begin{tabular}{lccc|ccc}
\multicolumn{4}{c|}{PDB ID:1AOH}&   \multicolumn{3}{c}{PDB ID:1TIT}\\ \hline
\diagbox[width=14em]{Model}{\hspace{1cm}Nr. of bonds/contacts} & EN & cm & Eff. Harm. (LJ) & EN & cm & Eff. Harm. (LJ)\\
\hline
Elastic Network          &   1131     &  -- & --   & 632   & --   & --  \\ 
GEN        &    352     & 349 & 430  & 203   & 156  & 273 \\
M1          &    352     & 779 & --   & 203   & 429  & --  \\
M2          &    352     & 349 & --   & 203   & 156  & --  \\
M3          &    782     & 349 & --   & 476   & 156  & --  \\
\end{tabular}
\end{table*}

 
 \begin{figure}[ht]
 \centering
   \includegraphics[scale =0.45]{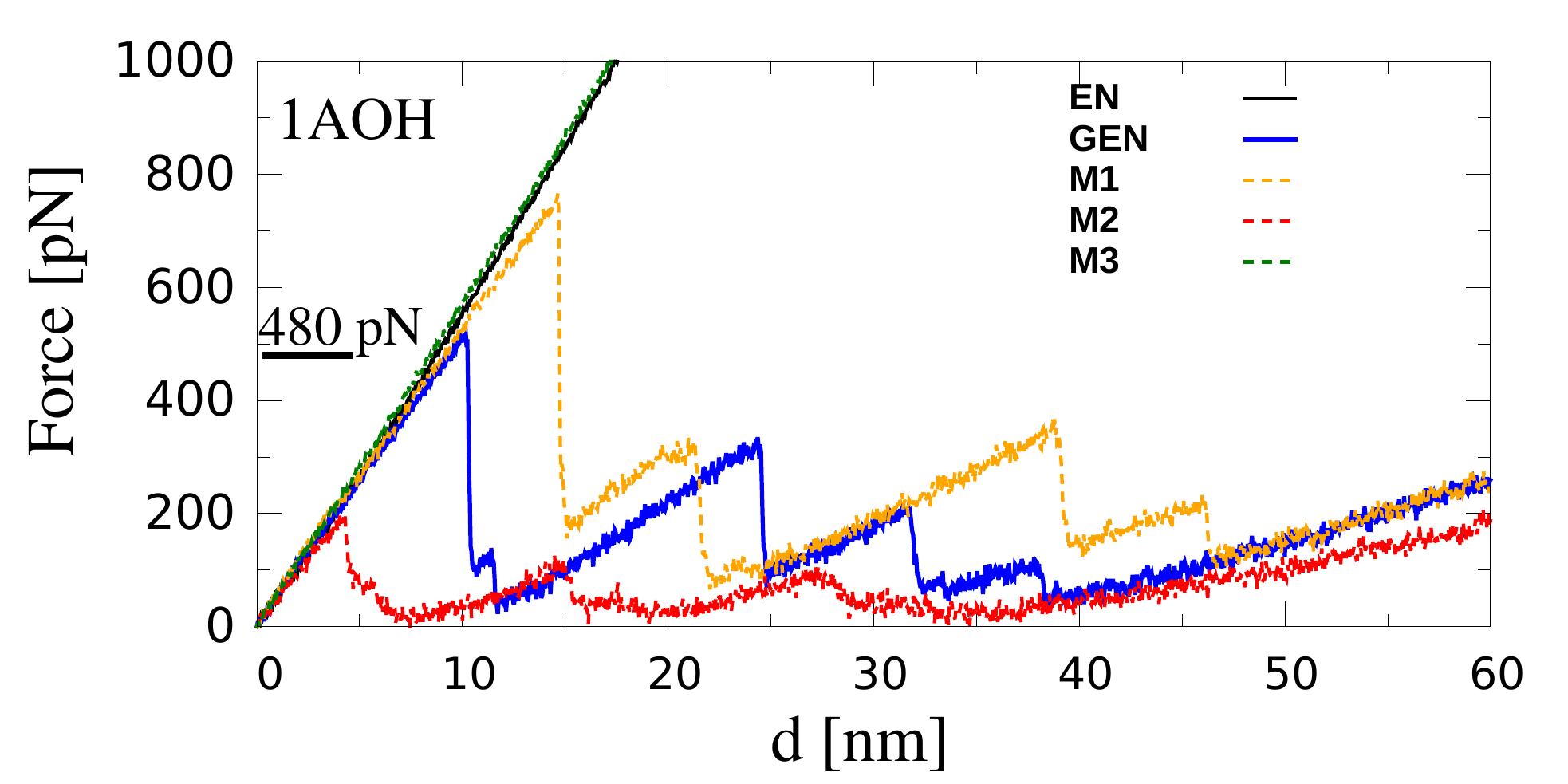}
   \includegraphics[scale =0.125]{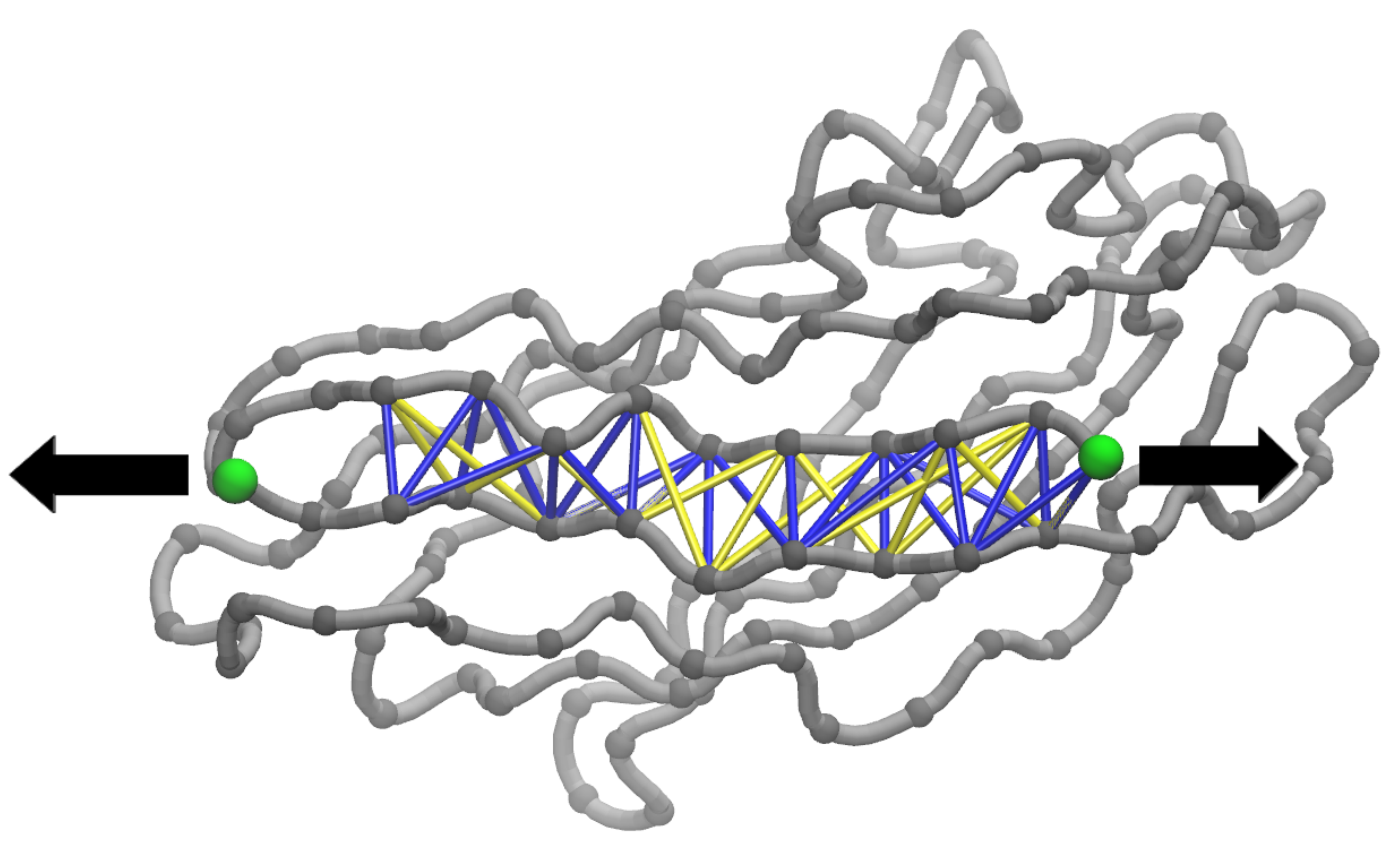}\hspace{2cm}
     \includegraphics[scale =0.15]{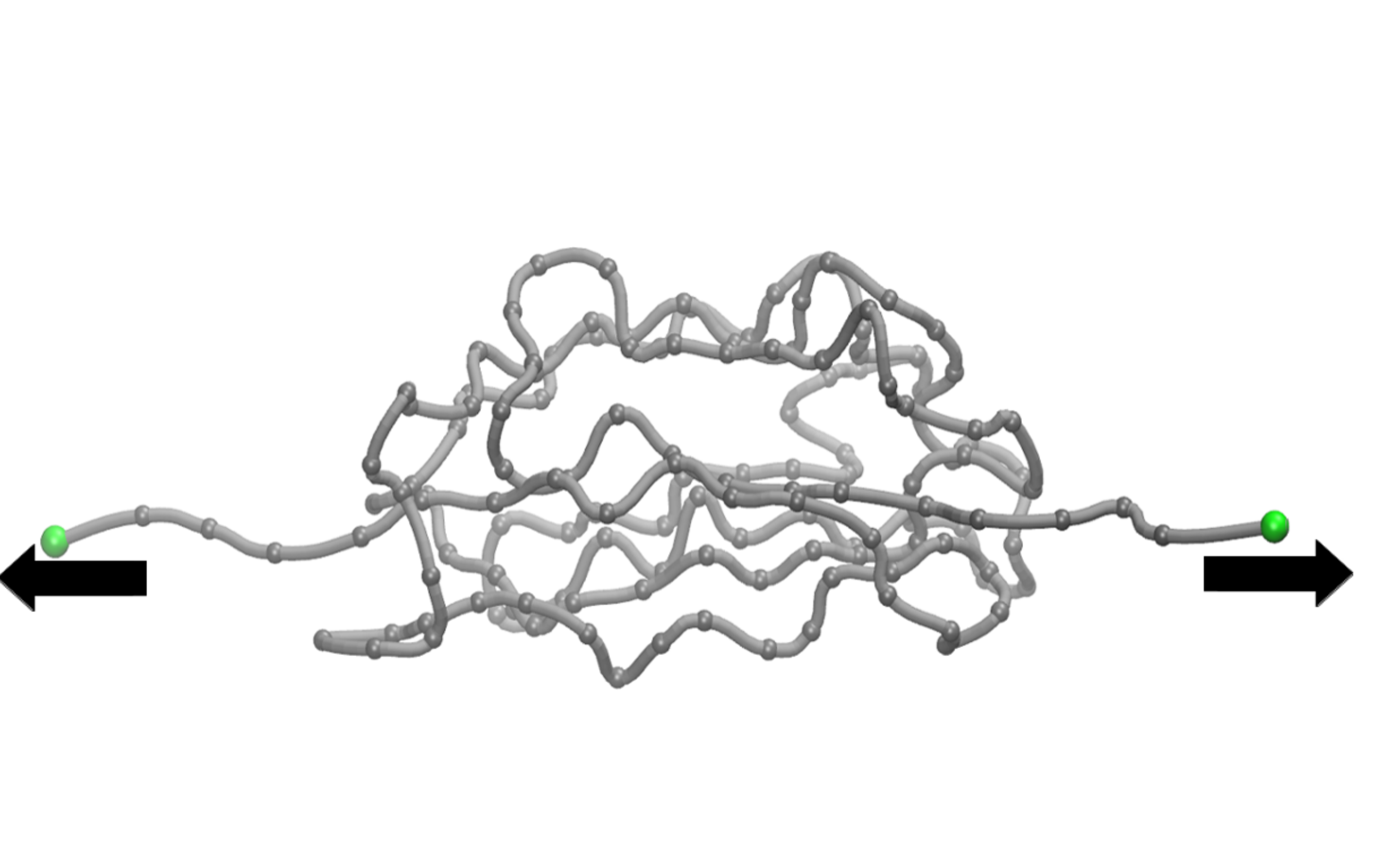}
   \caption{\small{Top panel shows the plot of force vs. cantilever displacement, $d$, for the type I cohesin domain (PDB ID: 1AOH). 
   Also, the experimental value for the maximum force, which is 480$\pm$14 pN \cite{Valbuena2009} in the case of 1AOH is indicated by a horizontal line. 
The bottom panel shows two snapshots at d = 0 nm and d = 10 nm. In the middle, we highlight the contacts responsible for the first peak (in blue) in GEN model and the additional yellow contacts that are present in the M1 model. We have performed a similar analysis for the I27 domain of titin (see SI). }}
   \label{pulling}
 \end{figure}

To further validate the GEN model, we have compared it with the standard EN and other possible versions of a breakable EN model (M1, M2 and M3)  for a number of
different protein structures determined by X-Ray diffraction and within 1-2 \AA~ of resolution. For this purpose, we performed  extensive  tests  on  the following proteins: The first one is about 124 residues (PDB ID: 5RSA) and corresponds to ribonuclease A\cite{wlodawer1986comparison}, the second one is obtained from the bovine pancreatic trypsin inhibitor\cite{wlodawer1984structure} with a length of 58 residues (PDB ID: 5PTI), and the last one corresponds to a  muscle  protein (G-actine) \cite{kabsch1990atomic} with 373 residues ((PDB ID: 1ATN).  The first two protein chains are made by helices and beta-strands while the last chain is folded so as to form two large domains joined by a narrow neck region. These  two  domains  are  partly  held  together  by salt bridges and hydrogen bonds provided by a nucleotide that stabilizes the two domains. Our NMA results are presented in Fig.~\ref{betafactor_5RSA_5PTI} along with their corresponding frequency data.
Clearly, the GEN model exhibits the best agreement with the EN, again
indicating the very good approximation of our harmonic terms with appropriate
effective LJ interactions and
the small influence of the native LJ contacts in the model. We have also checked
a number of additional proteins and we have found consistent results and a similar 
agreement between the GEN and the EN models. Moreover, the M3 model, which has undergone
a small modification by including the native LJ contacts in the EN, exhibits obviously almost
absolute agreement with the EN, whereas the M1 and M2 models show
considerable deviation from the EN, due to the lack of a large number of harmonic or effective-harmonic interactions [$U_{\rm LJ}(\epsilon_{\rm harm})$] (see Table~\ref{table1}).
This shows that the $U_{\rm LJ}(\epsilon_{\rm cm})$ terms are not enough to preserve the structure and properties of the targeted proteins without assuming extra
terms that contribute to backbone stiffness (\textit{e.g.} bond and dihedral angles). Moreover, the latter terms require tuning, as in the case of G\=o-like models. For a comparison of 64 different G\=o models, see Ref. \cite{64models}.

\subsection{Pullling simulations}

As our aim here is to propose an as simple and accurate as possible model for studying mechanical unfolding of
proteins, we have carried out pulling simulations in the same manner as in the case of single molecule studies performed with AFM \cite{Rief1997,Carrion-Vazquez1999}. We used
implicit solvent conditions similar to ref. \cite{Cieplak2002}. 
Overall, the correct redistribution of contacts between the above three categories 
(see Table \ref{table1}) results in the excellent agreement of our simulations results (GEN model)
with the experimental maximum pulling force, $F_{\rm max}$, 
in pulling simulations as is shown here for two examples, 
cohesin (PDB ID: 1AOH)  (Fig.~\ref{pulling}) and 
I27 domain of titin (PDB ID: 1TIT) (see Fig. S1 in SI).
The temperature during pulling simulations is 0.3$\epsilon_{\rm cm}/k_{\rm B}$.
The early unfolding scenario at the experimental pulling speed, which gives rise to the assessment of the mechanical properties of proteins,
is difficult to achieve by using all-atom simulations, because the time-scale involved is too short 
for stretching proteins in the case of all-atom methods. In particular, the typical speed used to stretch proteins in all-atom simulations is 
nowadays of the order of $10^{-2}$ nm/ps \cite{sotomayor2007single} and the experimental cantilever speed is around $10^{-9}$ nm/ps \cite{Carrion-Vazquez1999}. 
In this regard, the CG nature of our approach can be used to study a range of speeds much closer to the experimental conditions in comparison with all-atom models. 

Multiple proteins are linked sequentially and one can typically observe a number of corresponding peaks, 
which signal the full unfolding of individual protein modules. Due to the space resolution, 
intermediate unfolding states are not detected in AFM experiments \cite{Marszalek1999}. However, 
by using CG models one can usually access these intermediate states with a 
better resolution and assign to each of them a force peak \cite{Poma2017}. The largest of these force peaks, 
F$_{\rm max}$, defines the characteristic unfolding force for the whole protein domain. 

The GEN model is the best to reproduce the experimental rupture force for cohesin and the I27 domain of titin
(Fig.~\ref{pulling}). In particular, the maximum force is 480$\pm$14 and 204$\pm$30 pN for 1AOH, 
and 1TIT, respectively (see SI). The M2 model provided a much lower force peak, while
 a much higher one was observed in the case of the M1 model. This can be explained in terms of the number of contacts associated with the LJ interactions (see Fig.~\ref{pulling}, bottom panel), which in the case of M2 model appears to be smaller, but in the case of M1 is
much larger (Table~\ref{table1}). In addition, in the case of M3 and EN models there is no peak
due to the presence of the harmonic bonds between the C$_{\rm \alpha}$ atoms that prevent  the unfolding even at very large stretching forces. The unfolding pathway for 1AOH protein has been previously characterized by a G\=o-like model \cite{wojciechowski2014protein} and experiment \cite{valbuena2009remarkable}. It is known that the detachment of  $\beta_{1}(6-15)$ from $\beta_{9}(136-147)$ domains occurs at the same position of the maximum force in the $F$-$d$ plot. We have carried out the analysis of native contacts between pairs of $\beta$-strands which are responsible for stabilizing the protein (see SI). Our results capture the sequential detachment of the secondary structures that give rise to the largest force peak. Our observation agrees well with the breaking of native contacts between $\beta_{1}$ and $\beta_{9}$ strands. The characterization of the unfolding pathway for titin also agrees with the experimental results \cite{fowler2002mechanical,best2003mechanical} and is shown in SI. Moreover, proteins do not show any spurious effects with respect to their structure (\textit{e.g.} local aggregation) during the stretching due to the presence of harmonic bonds in the structure for our GEN model. 
We have further confirmed our conclusions by investigating a number of
different proteins.
 \begin{figure}[ht!]
	\centering
	\includegraphics[scale =0.45]{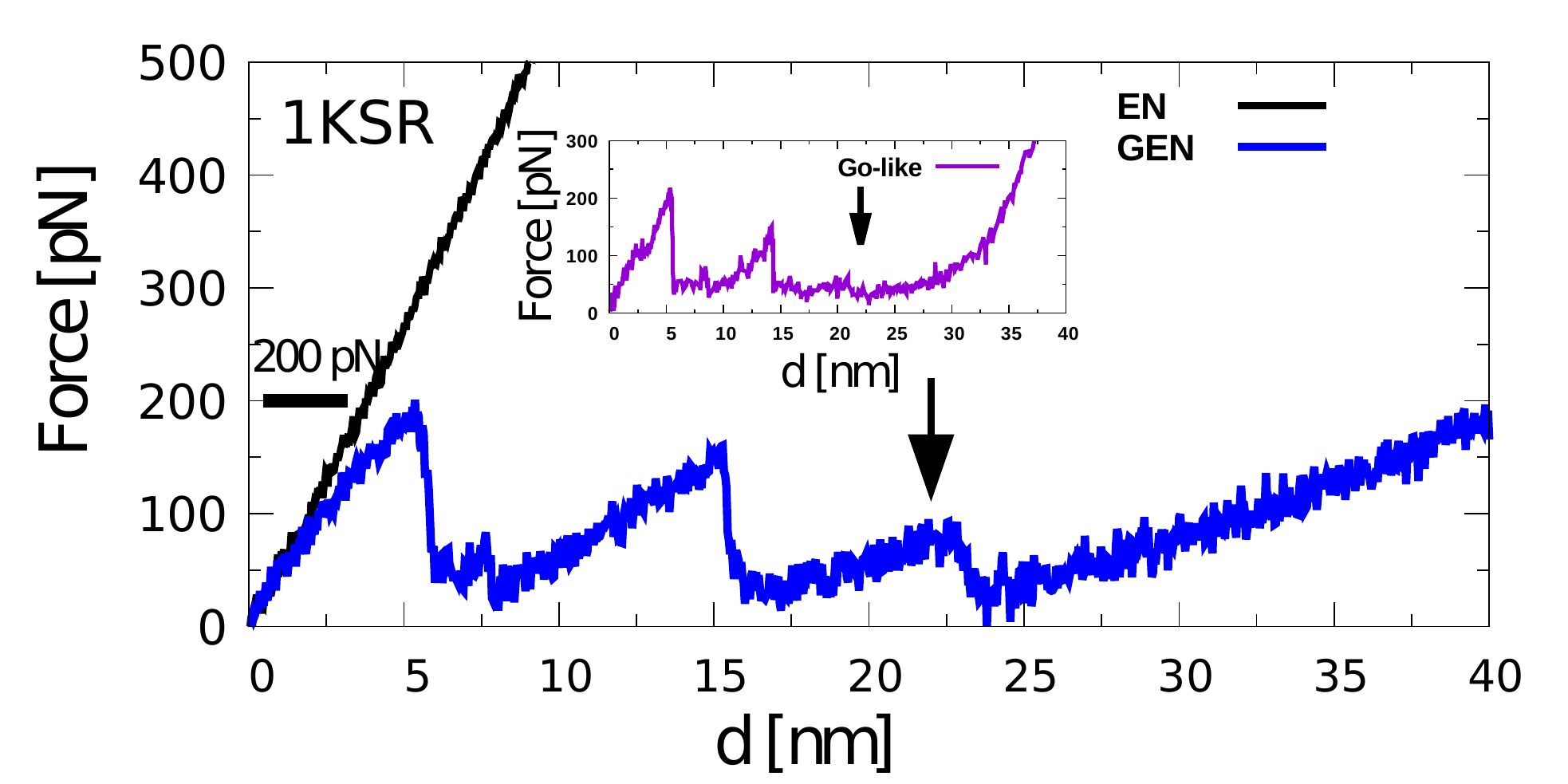}
	\includegraphics[scale =0.16]{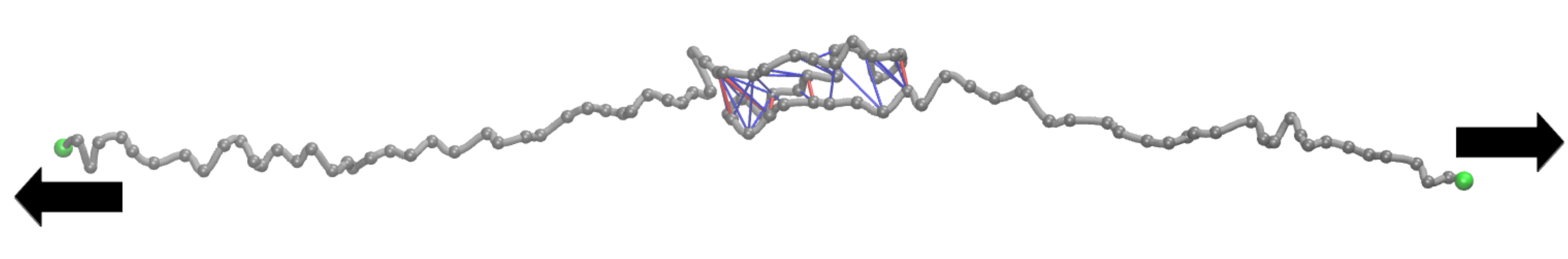}
	
	\caption{\small{Top panel shows the force--displacement profiles for domain 4 of the DDFLN4 protein. The results were obtained by using the GEN model (blue line), the EN model (black line) and the G\=o model (inset) for pulling speed v=$5\times10^{-4}$ nm/$\tau$, where $\tau$ is approximately 1ns. Results were averaged over 40 trajectories. Arrows refer to positions of the second peak at $d=22$ nm, which is expected to be the same as in experiments \cite{schwaiger2005folding,schwaiger2004mechanical}. Bottom panel illustrates a simulation snapshot for $d=22$ nm, where the contacts that stabilize the structures are in blue (G\=o-like) and in red (harmonic). The N- and C- termini beads are shown in green and the pulling direction is denoted by arrows. The data points for the G\=o-like model were extracted by using g3data software \cite{frantz2009g3data} from ref.\cite{sikora2009mechanical}. }}
	\label{1KSR}
\end{figure}

The mechanostability of the DDFLN4 protein (with PDB ID: 1KSR) has been studied experimentally \cite{schwaiger2005folding,schwaiger2004mechanical} and theoretically with all-atom simulation by Kouza \textit{et al.}. \cite{kouza2009protein}. In experiment, two peaks were observed in the force--displacement curve: one was located at $d=11$ nm and another one at $22$ nm. Here, we tested the performance of the GEN model against a standard G\=o-like model \cite{sikora2009mechanical} for this protein. The G\=o-like model captures the first experimental peak approximately at $d=13$ nm, but it misses the second peak (see Figure \ref{1KSR}). This is due to the lack of additional far distance contacts, which are only included by the GEN model (in GEN model these are treated by the harmonic approximation). In this regard, the GEN model performs better than the G\=o-like model as it reproduces both experimental peaks. Moreover, the GEN model is as good as the all-atom model \cite{kouza2009protein} in predicting mechanical unfolding intermediates because both models provide three peaks at nearly the same positions. Note that in our CG simulations we obtained an earlier force peak at $d=4$ nm, which is consistent with all-atom simulation \cite{kouza2009protein}. However, this peak has not been detected in experiment.

\subsection{Folding simulation of small peptides}

\begin{figure}[ht!]
	\centering
	\includegraphics[scale =0.4]{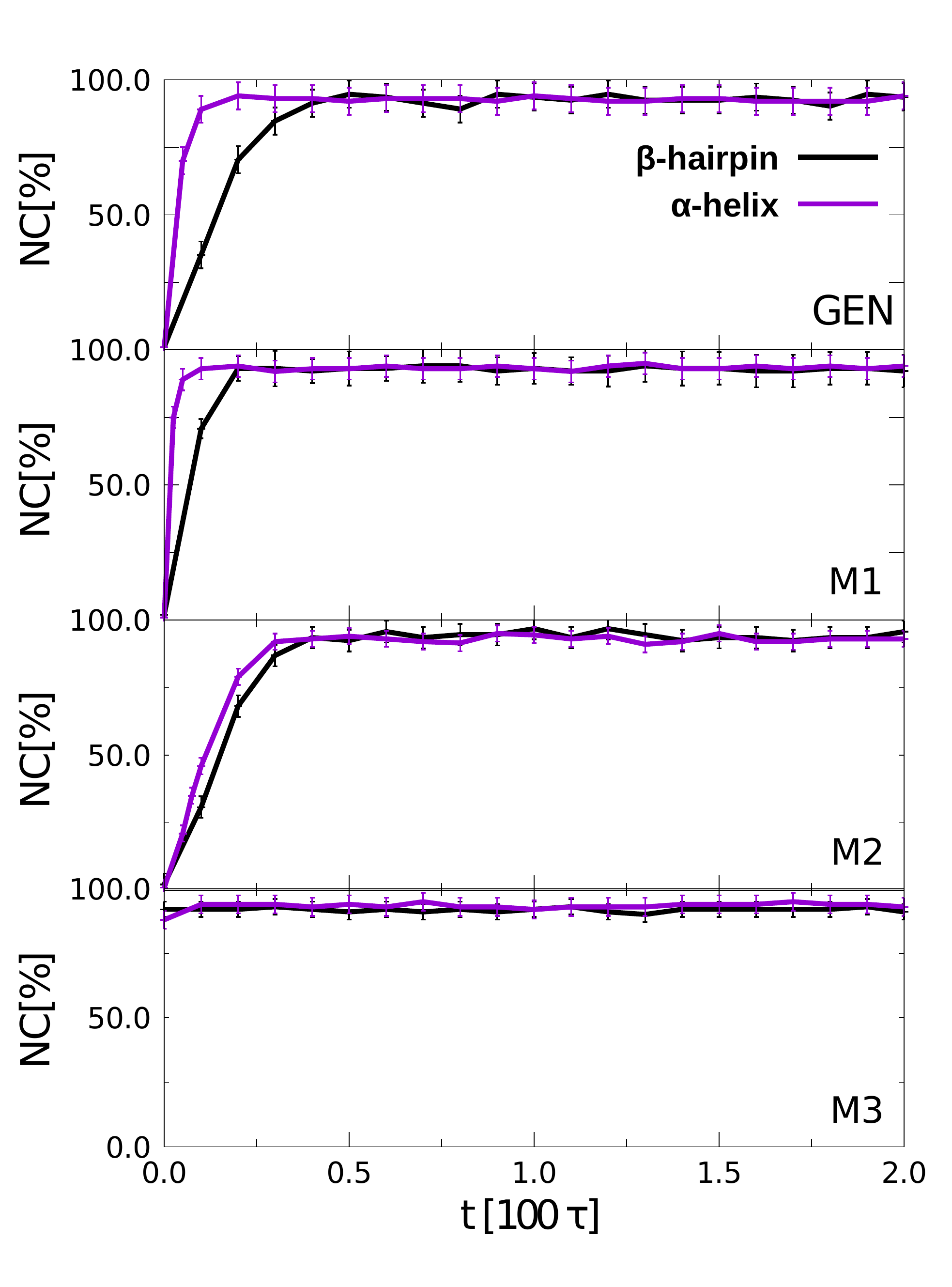}\\
	\includegraphics[scale =0.1]{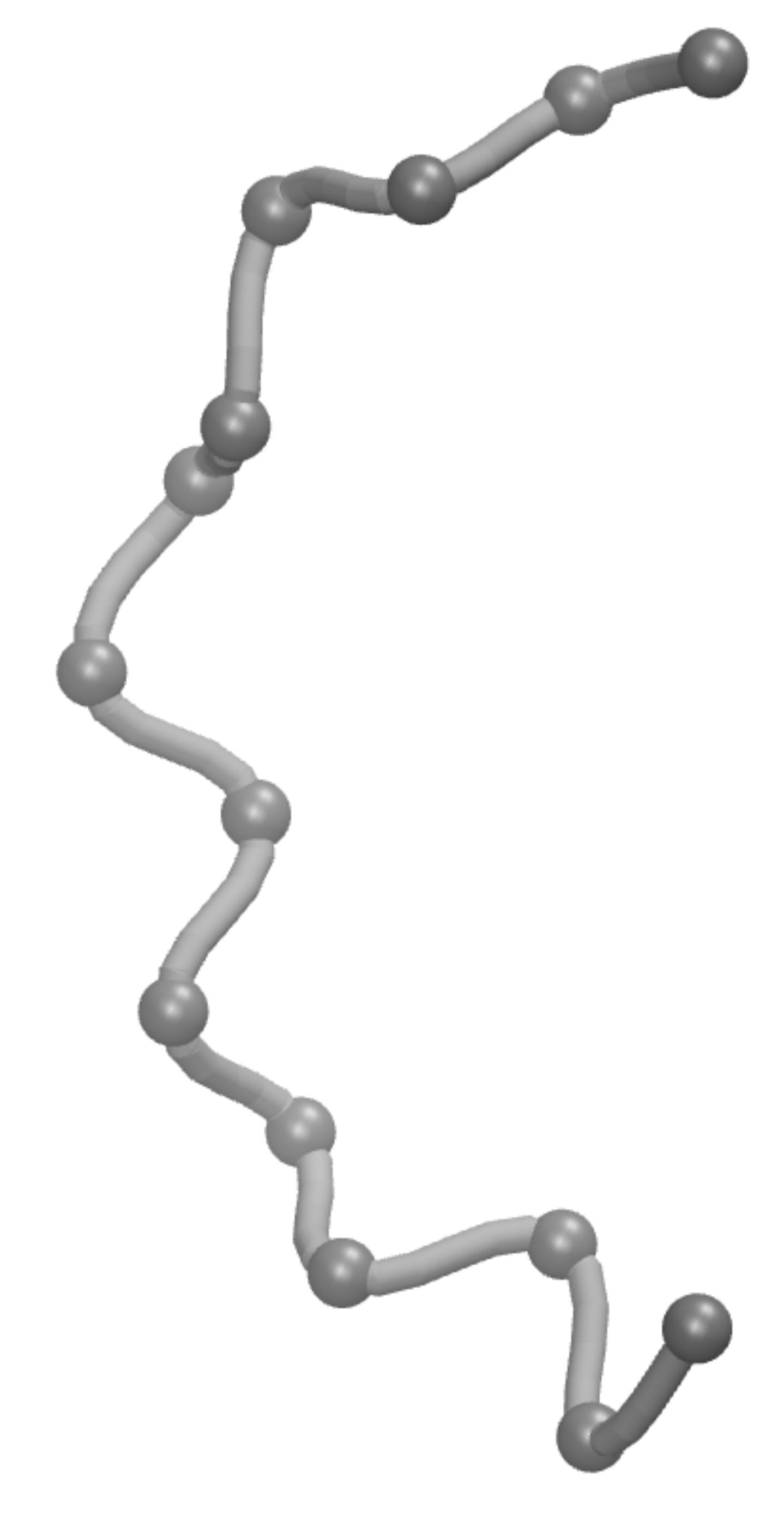}
	\includegraphics[scale =0.12]{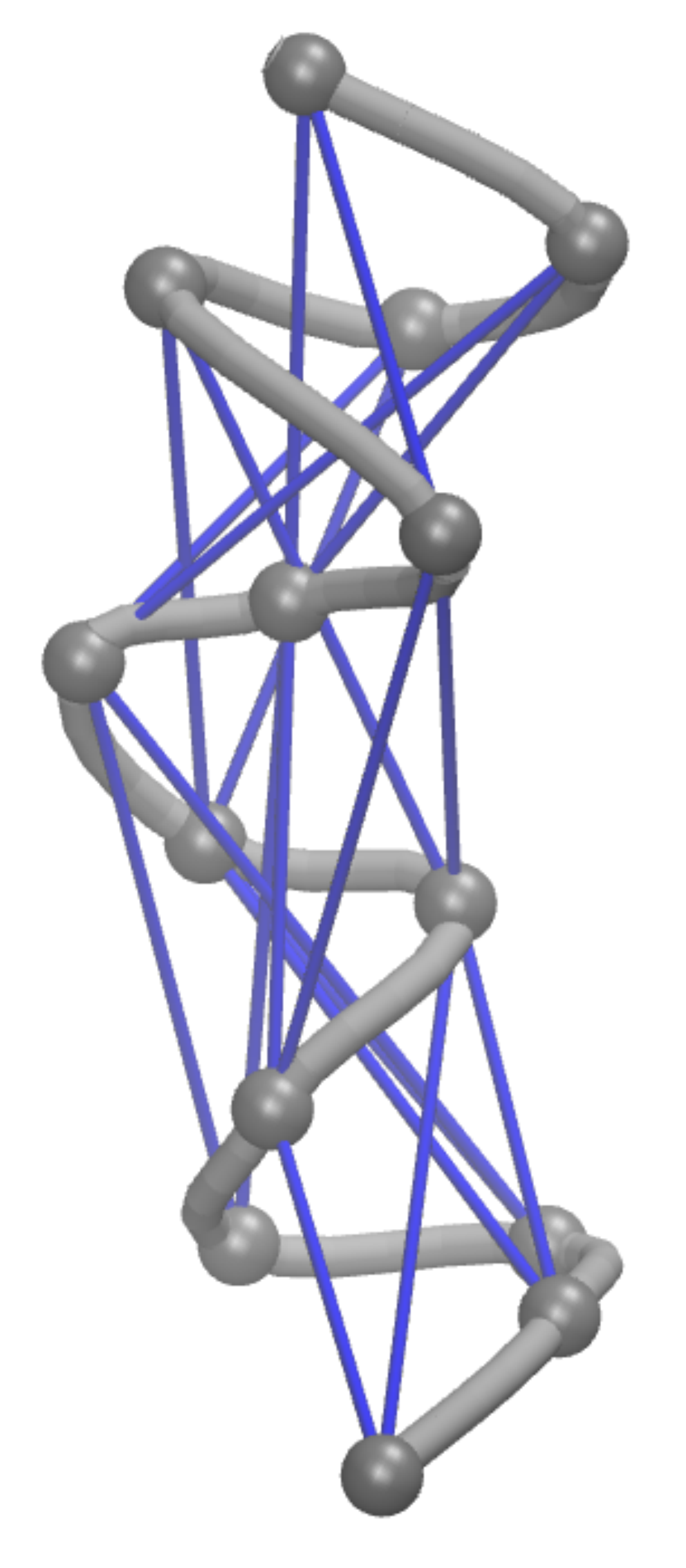}\hspace{2cm}
	\includegraphics[scale =0.1,angle=90]{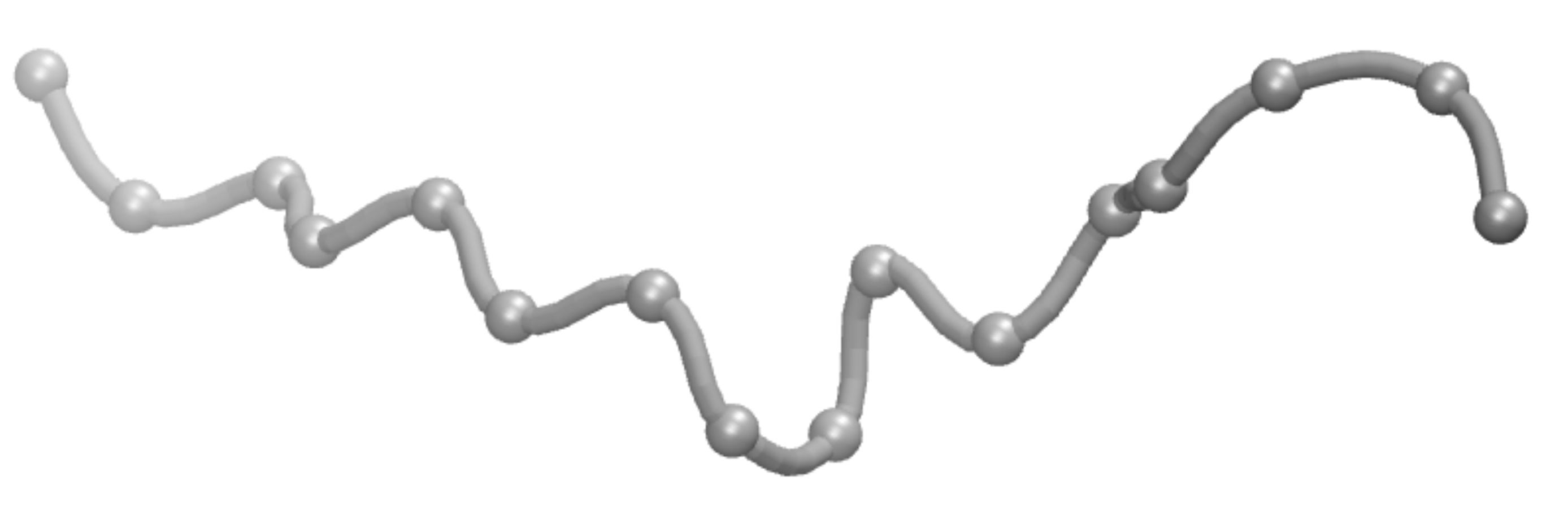}
	\includegraphics[scale =0.17]{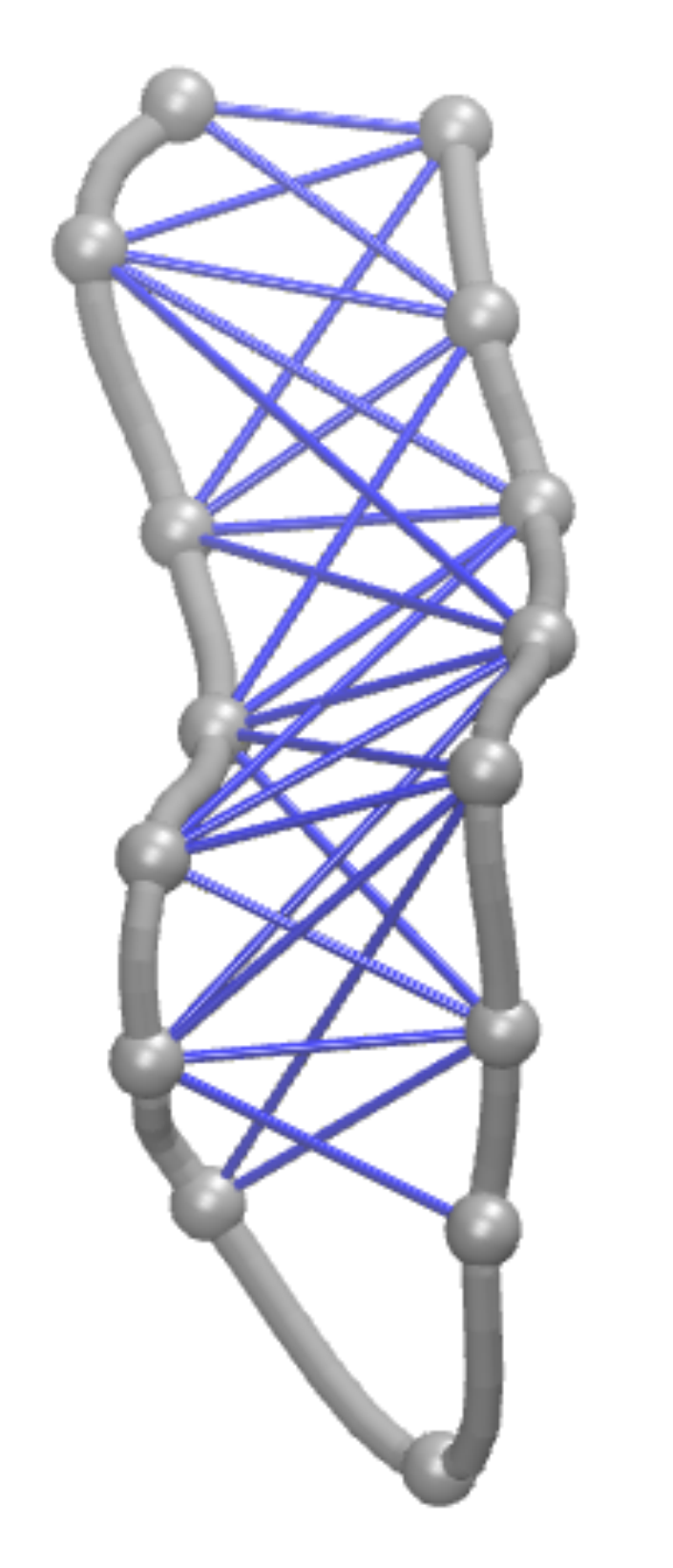}
	\caption{\small{Folding of two small peptides, a $\beta$-hairpin and an $\alpha$-helix. Plots show the percentage of G\=o-like and harmonic contacts present at a certain time during the folding process for each case and breakable model. Top panel for GEN model, middle panels for M1 model and M2 and bottom panel for M3 model, as indicated. A typical folding time for the $\alpha$-helix is about 10 $\tau$, while for the $\beta$-hairpin it is about 70 $\tau$ in the GEN model. Snapshots indicate examples of an unfolded state at the beginning of the simulation and a final folded (native) structure for each peptide. The C$_{\rm \alpha}$ atoms are represented by grey color. Native and harmonic contacts are described by solid blue lines.}}
	\label{folding}
\end{figure}

We are presenting here the folding process of two, well documented in the literature, small peptides, namely, an $\alpha$-helix comprising the sequence segment 70--83 of the protein HPr from Escherichia coli  (PDB ID: 1HDN \cite{van1994high} with 85 residues in total) and a $\beta$-hairpin (residues 41--56 of the immunoglobulin binding domain of streptococcal protein G with PDB ID: 1GB1 \cite{gronenborn1991novel} and 56 residues in total). Initial configurations for MD simulation are  unfolded conformations without any G\=o-like or effective harmonic contacts. According to a standard criterion that is commonly used in the case of G\=o-like models, the G\=o-like contacts are present in the structure when the actual distance between two C$_{\rm \alpha}$ atoms is smaller than 1.5$\sigma_{\rm ij}$, where $r_{\rm ij}^{0}$  is the distance between two C$_{\rm \alpha}$ atoms that form a contact in the native conformation. Unfolded
 structures were obtained by heating up the system at 500 K without water and 
making sure that no native contacts are present in the protein structure by using the above criterion based on the distance between the C$_{\rm \alpha}$ atoms in the native structure. 
In this way, we produced initial configurations
 for 50 statistically independent MD trajectories of length 200 $\tau$ at 300 K. Figure \ref{folding} shows the convergence of the total number of contacts towards the folded structure.

The GEN model and two of its variants (M1 and M2) allow for the study of protein folding, whereas the EN model apparently is not suitable for folding studies,
due to the presence of the harmonic bonds. Our results and typical snapshots of unfolded and well-folded (native) structures are presented in Figure \ref{folding}.
In the case of the $\alpha$-helix, the folding did occur in all independent trajectories, whereas about  5\% of the trajectories in the case of $\beta$-hairpin did not reach the
native structure within the time scale of the simulation. Assuming that the time unit $\tau$=1 ns, we obtained the folding times for $\beta$-hairpin and $\alpha$-helix equal
to $t_{fold}^{\beta}$= 70 ns  and
$t_{fold}^{\alpha}$= 10 ns , respectively. According to the temperature-jump fluorescence experiment by Munoz \textit{et al.} \cite{Munoz_Nature1997},  $t_{fold}^{\beta} \approx 6 \mu$s. Since the folding time of $\alpha$-helix is about 0.7 $\mu$s \cite{kubelka2004protein} the experimental ratio $t_{fold}^{\beta}/t_{fold}^{\alpha} \approx$ 9,  which is close to the value of 7, which was obtained from our simulations.  In this regard, the other models underestimate the ratio. For instance, M1 and M2 give approximately 2 and 1.3, respectively. Moreover, the M3 model does not capture any distinction between both peptides. This is due to the presence of the harmonic contacts that place the unfolded state in a very high energy state. This induces a rapid process towards the folded state just after energy minimization. Still, the absolute folding time in GEN is much shorter than the real time, due to the coarse-graining, while all-atom simulations from different groups have predicted values in the range 1--7 $\mu$s \cite{Phuong_Proteins2005,Zagrovic_JMB2001,Garcia_Proteins2001}. The fact that folding of the $\alpha$-helix is about seven times faster than the folding of the $\beta$-hairpin in the GEN model is also in agreement with our previous study \cite{Poma2017}, whereas M1 and M2 lead to a shorter time scale separation.


\section{\label{sec:summary} Conclusions}
In conclusion, this work presents a simple and apparently accurate model based on the EN approach for
studying large conformational changes in proteins.
Here, we have shown that the GEN model, despite its simplicity, maintains a close match with the EN, while
it reproduces with high accuracy the maximum force in AFM-pulling experiments and the folding time scales of peptides. On the one hand, there are several limitations in modeling proteins by CG approaches in particular due to the lack of details (\textit{e.g.,} solvent effect, amino acid specificity, \textit{etc.}) and thus we do not expect to capture all possible effects that stabilize protein complexes. However, our model handles native interactions with simplicity (G\=o-like potentials), which is crucial for enabling conformational changes. 
Moreover, the effective harmonic interactions described by LJ potentials and the native G\=o-like potentials prevent the steric clashes during the studies carried out in the present work. However, more sophisticated functional forms of non-native interactions could be included \textit{a posteriori} and their effect may be relevant in other applications.
The GEN model has enabled the study of protein folding confirming the timescale separation (about seven-fold difference in folding time between the $\alpha$-helix and the $\beta$-hairpin). 
On the other hand, our model uses a reduced number of parameters in comparison with any structured-based CG model that enables large conformational studies, while its foundation is based on the simple EN model with no assumptions about backbone connectivity. As we have shown, the GEN model provides the same number of peaks in the force--displacement profile as observed in the case of the all-atom models for the DDFLN4 protein. This result highlights the advantage of our model over standard Go-like models. In perspective, one can interface the GEN model with knowledge-based and free-energy derived potentials for the study of protein aggregation phenomena. It could also be used to study denaturation phenomena, for example, due to large changes in temperature or pressure. Such and other phenomena could possibly be described by our simple EN-type model and it would be interesting to check in the future the prediction power of GEN for different protein systems.

\section*{Acknowledgements}
This research has been supported by the National Science Centre, Poland, under grant No.~2015/19/P/ST3/03541, No.~2015/19/B/ST4/02721, and No.~2017/26/D/NZ1/00466. This project has received funding from the European Union's Horizon 2020 research and innovation programme under the Marie Sk{\l}odowska--Curie grant agreement No. 665778. This research was supported in part by PLGrid Infrastructure.

\balance


\bibliography{rsc} 

\providecommand*{\mcitethebibliography}{\thebibliography}
\csname @ifundefined\endcsname{endmcitethebibliography}
{\let\endmcitethebibliography\endthebibliography}{}
\begin{mcitethebibliography}{76}
\providecommand*{\natexlab}[1]{#1}
\providecommand*{\mciteSetBstSublistMode}[1]{}
\providecommand*{\mciteSetBstMaxWidthForm}[2]{}
\providecommand*{\mciteBstWouldAddEndPuncttrue}
  {\def\EndOfBibitem{\unskip.}}
\providecommand*{\mciteBstWouldAddEndPunctfalse}
  {\let\EndOfBibitem\relax}
\providecommand*{\mciteSetBstMidEndSepPunct}[3]{}
\providecommand*{\mciteSetBstSublistLabelBeginEnd}[3]{}
\providecommand*{\EndOfBibitem}{}
\mciteSetBstSublistMode{f}
\mciteSetBstMaxWidthForm{subitem}
{(\emph{\alph{mcitesubitemcount}})}
\mciteSetBstSublistLabelBeginEnd{\mcitemaxwidthsubitemform\space}
{\relax}{\relax}

\bibitem[Tirion(1996)]{Tirion1996}
M.~M. Tirion, \emph{Phys. Rev. Lett.}, 1996, \textbf{77}, 1905--1908\relax
\mciteBstWouldAddEndPuncttrue
\mciteSetBstMidEndSepPunct{\mcitedefaultmidpunct}
{\mcitedefaultendpunct}{\mcitedefaultseppunct}\relax
\EndOfBibitem
\bibitem[Setny and Zacharias(2013)]{Setny2013}
P.~Setny and M.~Zacharias, \emph{J. Chem. Theory Comput.}, 2013, \textbf{9},
  5460--5470\relax
\mciteBstWouldAddEndPuncttrue
\mciteSetBstMidEndSepPunct{\mcitedefaultmidpunct}
{\mcitedefaultendpunct}{\mcitedefaultseppunct}\relax
\EndOfBibitem
\bibitem[Zimmermann and Jernigan(2014)]{Zimmermann2014}
M.~T. Zimmermann and R.~L. Jernigan, \emph{RNA}, 2014, \textbf{20},
  792--804\relax
\mciteBstWouldAddEndPuncttrue
\mciteSetBstMidEndSepPunct{\mcitedefaultmidpunct}
{\mcitedefaultendpunct}{\mcitedefaultseppunct}\relax
\EndOfBibitem
\bibitem[Pinamonti \emph{et~al.}(2015)Pinamonti, Bottaro, Micheletti, and
  Bussi]{Pinamonti2015}
G.~Pinamonti, S.~Bottaro, C.~Micheletti and G.~Bussi, \emph{Nucleic Acids
  Res.}, 2015, \textbf{43}, 7260--7269\relax
\mciteBstWouldAddEndPuncttrue
\mciteSetBstMidEndSepPunct{\mcitedefaultmidpunct}
{\mcitedefaultendpunct}{\mcitedefaultseppunct}\relax
\EndOfBibitem
\bibitem[Kim \emph{et~al.}(2014)Kim, Kim, Choi, and Kim]{KiM2014}
M.~H. Kim, D.~Kim, J.~B. Choi and M.~K. Kim, \emph{Phys. Chem. Chem. Phys.},
  2014, \textbf{16}, 15263--15271\relax
\mciteBstWouldAddEndPuncttrue
\mciteSetBstMidEndSepPunct{\mcitedefaultmidpunct}
{\mcitedefaultendpunct}{\mcitedefaultseppunct}\relax
\EndOfBibitem
\bibitem[Glass \emph{et~al.}(2012)Glass, Moritsugu, Cheng, and
  Smith]{Glass2012}
D.~C. Glass, K.~Moritsugu, X.~Cheng and J.~C. Smith, \emph{Biomacromolecules},
  2012, \textbf{13}, 2634--2644\relax
\mciteBstWouldAddEndPuncttrue
\mciteSetBstMidEndSepPunct{\mcitedefaultmidpunct}
{\mcitedefaultendpunct}{\mcitedefaultseppunct}\relax
\EndOfBibitem
\bibitem[Cui and Bahar(2006)]{Cui2006}
Q.~Cui and I.~Bahar, \emph{Normal Mode Analysis. Theory and applications to
  biological and chemical systems.}, Chapman \& Hall/CRC, 2006\relax
\mciteBstWouldAddEndPuncttrue
\mciteSetBstMidEndSepPunct{\mcitedefaultmidpunct}
{\mcitedefaultendpunct}{\mcitedefaultseppunct}\relax
\EndOfBibitem
\bibitem[Bahar \emph{et~al.}(1997)Bahar, Atilgan, and Erman]{Bahar1997}
I.~Bahar, A.~R. Atilgan and B.~Erman, \emph{Fold. Design}, 1997, \textbf{2},
  173--181\relax
\mciteBstWouldAddEndPuncttrue
\mciteSetBstMidEndSepPunct{\mcitedefaultmidpunct}
{\mcitedefaultendpunct}{\mcitedefaultseppunct}\relax
\EndOfBibitem
\bibitem[Haliloglu \emph{et~al.}(1997)Haliloglu, Bahar, and Erman]{Bahar_PRL}
T.~Haliloglu, I.~Bahar and B.~Erman, \emph{Phys. Rev. Lett.}, 1997,
  \textbf{79}, 3090\relax
\mciteBstWouldAddEndPuncttrue
\mciteSetBstMidEndSepPunct{\mcitedefaultmidpunct}
{\mcitedefaultendpunct}{\mcitedefaultseppunct}\relax
\EndOfBibitem
\bibitem[Hinsen(1998)]{Hinsen1998}
K.~Hinsen, \emph{Proteins}, 1998, \textbf{33}, 417--429\relax
\mciteBstWouldAddEndPuncttrue
\mciteSetBstMidEndSepPunct{\mcitedefaultmidpunct}
{\mcitedefaultendpunct}{\mcitedefaultseppunct}\relax
\EndOfBibitem
\bibitem[Hinsen and Field(1999)]{Hinsen1999}
A.~Hinsen, K.~Thomas and M.~Field, \emph{Proteins}, 1999, \textbf{34},
  369--382\relax
\mciteBstWouldAddEndPuncttrue
\mciteSetBstMidEndSepPunct{\mcitedefaultmidpunct}
{\mcitedefaultendpunct}{\mcitedefaultseppunct}\relax
\EndOfBibitem
\bibitem[Atilgan \emph{et~al.}(2001)Atilgan, Durell, Jernigan, Demirel, Keskin,
  and Bahar]{Atilgan2001}
A.~R. Atilgan, S.~R. Durell, R.~L. Jernigan, M.~C. Demirel, O.~Keskin and
  I.~Bahar, \emph{Biophys. J.}, 2001, \textbf{80}, 505--515\relax
\mciteBstWouldAddEndPuncttrue
\mciteSetBstMidEndSepPunct{\mcitedefaultmidpunct}
{\mcitedefaultendpunct}{\mcitedefaultseppunct}\relax
\EndOfBibitem
\bibitem[Tama and Sanejouand(2001)]{Tama2001}
F.~Tama and Y.-H. Sanejouand, \emph{Protein Eng.}, 2001, \textbf{14}, 1\relax
\mciteBstWouldAddEndPuncttrue
\mciteSetBstMidEndSepPunct{\mcitedefaultmidpunct}
{\mcitedefaultendpunct}{\mcitedefaultseppunct}\relax
\EndOfBibitem
\bibitem[Kim \emph{et~al.}(2002)Kim, Jernigan, and
  Chirikjian]{kiM2002efficient}
M.~K. Kim, R.~L. Jernigan and G.~S. Chirikjian, \emph{Biophys. J.}, 2002,
  \textbf{83}, 1620--1630\relax
\mciteBstWouldAddEndPuncttrue
\mciteSetBstMidEndSepPunct{\mcitedefaultmidpunct}
{\mcitedefaultendpunct}{\mcitedefaultseppunct}\relax
\EndOfBibitem
\bibitem[Feng \emph{et~al.}(2009)Feng, Yang, Kloczkowski, and
  Jernigan]{feng2009energy}
Y.~Feng, L.~Yang, A.~Kloczkowski and R.~L. Jernigan, \emph{Proteins: Struct.,
  Funct., Bioinf.}, 2009, \textbf{77}, 551--558\relax
\mciteBstWouldAddEndPuncttrue
\mciteSetBstMidEndSepPunct{\mcitedefaultmidpunct}
{\mcitedefaultendpunct}{\mcitedefaultseppunct}\relax
\EndOfBibitem
\bibitem[Das \emph{et~al.}(2014)Das, Gur, Cheng, Jo, Bahar, and
  Roux]{das2014exploring}
A.~Das, M.~Gur, M.~H. Cheng, S.~Jo, I.~Bahar and B.~Roux, \emph{PLOS comput.
  Biol.}, 2014, \textbf{10}, e1003521\relax
\mciteBstWouldAddEndPuncttrue
\mciteSetBstMidEndSepPunct{\mcitedefaultmidpunct}
{\mcitedefaultendpunct}{\mcitedefaultseppunct}\relax
\EndOfBibitem
\bibitem[Tekpinar and Zheng(2010)]{tekpinar2010predicting}
M.~Tekpinar and W.~Zheng, \emph{Proteins: Struct., Funct., Bioinf.}, 2010,
  \textbf{78}, 2469--2481\relax
\mciteBstWouldAddEndPuncttrue
\mciteSetBstMidEndSepPunct{\mcitedefaultmidpunct}
{\mcitedefaultendpunct}{\mcitedefaultseppunct}\relax
\EndOfBibitem
\bibitem[Clementi \emph{et~al.}(2000)Clementi, Nymeyer, and
  Onuchic]{Clementi2000}
C.~Clementi, H.~Nymeyer and J.~N. Onuchic, \emph{J. Mol. Biol.}, 2000,
  \textbf{298}, 937--953\relax
\mciteBstWouldAddEndPuncttrue
\mciteSetBstMidEndSepPunct{\mcitedefaultmidpunct}
{\mcitedefaultendpunct}{\mcitedefaultseppunct}\relax
\EndOfBibitem
\bibitem[Karanicolas and Brooks(2002)]{Karanicolas2002}
J.~Karanicolas and C.~L. Brooks, \emph{Protein Sci.}, 2002, \textbf{11},
  2351--2361\relax
\mciteBstWouldAddEndPuncttrue
\mciteSetBstMidEndSepPunct{\mcitedefaultmidpunct}
{\mcitedefaultendpunct}{\mcitedefaultseppunct}\relax
\EndOfBibitem
\bibitem[Poma \emph{et~al.}(2017)Poma, Cieplak, and Theodorakis]{Poma2017}
A.~B. Poma, M.~Cieplak and P.~E. Theodorakis, \emph{J. Chem. Theory Comput.},
  2017, \textbf{13}, 1366--1374\relax
\mciteBstWouldAddEndPuncttrue
\mciteSetBstMidEndSepPunct{\mcitedefaultmidpunct}
{\mcitedefaultendpunct}{\mcitedefaultseppunct}\relax
\EndOfBibitem
\bibitem[Van~Wynsberghe \emph{et~al.}(2004)Van~Wynsberghe, Li, and
  Cui]{van2004normal}
A.~Van~Wynsberghe, G.~Li and Q.~Cui, \emph{Biochemistry}, 2004, \textbf{43},
  13083--13096\relax
\mciteBstWouldAddEndPuncttrue
\mciteSetBstMidEndSepPunct{\mcitedefaultmidpunct}
{\mcitedefaultendpunct}{\mcitedefaultseppunct}\relax
\EndOfBibitem
\bibitem[Bagci \emph{et~al.}(2003)Bagci, Kloczkowski, Jernigan, and
  Bahar]{bagci2003origin}
Z.~Bagci, A.~Kloczkowski, R.~L. Jernigan and I.~Bahar, \emph{Proteins: Struct.,
  Funct., Bioinf.}, 2003, \textbf{53}, 56--67\relax
\mciteBstWouldAddEndPuncttrue
\mciteSetBstMidEndSepPunct{\mcitedefaultmidpunct}
{\mcitedefaultendpunct}{\mcitedefaultseppunct}\relax
\EndOfBibitem
\bibitem[Fletcher and Powell(1963)]{fletcher1963rapidly}
R.~Fletcher and M.~J. Powell, \emph{Comput. J.}, 1963, \textbf{6},
  163--168\relax
\mciteBstWouldAddEndPuncttrue
\mciteSetBstMidEndSepPunct{\mcitedefaultmidpunct}
{\mcitedefaultendpunct}{\mcitedefaultseppunct}\relax
\EndOfBibitem
\bibitem[Kershaw(1978)]{kershaw1978incomplete}
D.~S. Kershaw, \emph{J. Comput. Phys.}, 1978, \textbf{26}, 43--65\relax
\mciteBstWouldAddEndPuncttrue
\mciteSetBstMidEndSepPunct{\mcitedefaultmidpunct}
{\mcitedefaultendpunct}{\mcitedefaultseppunct}\relax
\EndOfBibitem
\bibitem[Periole \emph{et~al.}(2009)Periole, Cavalli, Marrink, and
  Ceruso]{periole2009combining}
X.~Periole, M.~Cavalli, S.-J. Marrink and M.~A. Ceruso, \emph{J. Chem. Theory
  Comput.}, 2009, \textbf{5}, 2531--2543\relax
\mciteBstWouldAddEndPuncttrue
\mciteSetBstMidEndSepPunct{\mcitedefaultmidpunct}
{\mcitedefaultendpunct}{\mcitedefaultseppunct}\relax
\EndOfBibitem
\bibitem[Rief \emph{et~al.}(1997)Rief, Gautel, Oesterhelt, Fernandez, and
  Gaub]{Rief1997}
M.~Rief, M.~Gautel, F.~Oesterhelt, J.~M. Fernandez and H.~E. Gaub,
  \emph{Science}, 1997, \textbf{276}, 1109--1112\relax
\mciteBstWouldAddEndPuncttrue
\mciteSetBstMidEndSepPunct{\mcitedefaultmidpunct}
{\mcitedefaultendpunct}{\mcitedefaultseppunct}\relax
\EndOfBibitem
\bibitem[Kellermayer \emph{et~al.}(1997)Kellermayer, Smith, Granzier, and
  Bustamante]{Kellermayer1997}
M.~S. Kellermayer, S.~B. Smith, H.~L. Granzier and C.~Bustamante,
  \emph{Science}, 1997, \textbf{276}, 1112--1116\relax
\mciteBstWouldAddEndPuncttrue
\mciteSetBstMidEndSepPunct{\mcitedefaultmidpunct}
{\mcitedefaultendpunct}{\mcitedefaultseppunct}\relax
\EndOfBibitem
\bibitem[Su\l{}kowska \emph{et~al.}(2008)Su\l{}kowska, Kloczkowski, Sen,
  Cieplak, and Jernigan]{Sulkowska2008}
J.~Su\l{}kowska, A.~Kloczkowski, T.~Sen, M.~Cieplak and R.~Jernigan,
  \emph{Proteins}, 2008, \textbf{71}, 45--60\relax
\mciteBstWouldAddEndPuncttrue
\mciteSetBstMidEndSepPunct{\mcitedefaultmidpunct}
{\mcitedefaultendpunct}{\mcitedefaultseppunct}\relax
\EndOfBibitem
\bibitem[Kumar and Li(2010)]{Kumar2010}
S.~Kumar and M.~S. Li, \emph{Phys. Rep.}, 2010, \textbf{486}, 1 -- 74\relax
\mciteBstWouldAddEndPuncttrue
\mciteSetBstMidEndSepPunct{\mcitedefaultmidpunct}
{\mcitedefaultendpunct}{\mcitedefaultseppunct}\relax
\EndOfBibitem
\bibitem[Becker \emph{et~al.}(2003)Becker, Oroudjev, Mutz, Cleveland, Hansma,
  Hayashi, Makarov, and Hansma]{Becker2003}
N.~Becker, E.~Oroudjev, S.~Mutz, J.~P. Cleveland, P.~K. Hansma, C.~Y. Hayashi,
  D.~E. Makarov and H.~G. Hansma, \emph{Nat. Mater.}, 2003, \textbf{2},
  278--283\relax
\mciteBstWouldAddEndPuncttrue
\mciteSetBstMidEndSepPunct{\mcitedefaultmidpunct}
{\mcitedefaultendpunct}{\mcitedefaultseppunct}\relax
\EndOfBibitem
\bibitem[Landau and Lifshitz(1986)]{Landau1986}
L.~D. Landau and E.~Lifshitz, \emph{Course of Theoretical Physics}, 1986,
  \textbf{3}, 109\relax
\mciteBstWouldAddEndPuncttrue
\mciteSetBstMidEndSepPunct{\mcitedefaultmidpunct}
{\mcitedefaultendpunct}{\mcitedefaultseppunct}\relax
\EndOfBibitem
\bibitem[Jackson(1998)]{Jackson1998}
S.~E. Jackson, \emph{Fold. Des.}, 1998, \textbf{3}, R81--R91\relax
\mciteBstWouldAddEndPuncttrue
\mciteSetBstMidEndSepPunct{\mcitedefaultmidpunct}
{\mcitedefaultendpunct}{\mcitedefaultseppunct}\relax
\EndOfBibitem
\bibitem[Benjwal \emph{et~al.}(2006)Benjwal, Verma, R{\"o}hm, and
  Gursky]{Benjwal2006}
S.~Benjwal, S.~Verma, K.-H. R{\"o}hm and O.~Gursky, \emph{Protein Sci.}, 2006,
  \textbf{15}, 635--639\relax
\mciteBstWouldAddEndPuncttrue
\mciteSetBstMidEndSepPunct{\mcitedefaultmidpunct}
{\mcitedefaultendpunct}{\mcitedefaultseppunct}\relax
\EndOfBibitem
\bibitem[Hillson \emph{et~al.}(1999)Hillson, Onuchic, and
  Garc{\'\i}a]{pressure}
N.~Hillson, J.~N. Onuchic and A.~E. Garc{\'\i}a, \emph{Proc. Natl. Acad. Sci.
  U.S.A.}, 1999, \textbf{96}, 14848--14853\relax
\mciteBstWouldAddEndPuncttrue
\mciteSetBstMidEndSepPunct{\mcitedefaultmidpunct}
{\mcitedefaultendpunct}{\mcitedefaultseppunct}\relax
\EndOfBibitem
\bibitem[Otzen(2002)]{surfactant}
D.~E. Otzen, \emph{Biophys. J.}, 2002, \textbf{83}, 2219--2230\relax
\mciteBstWouldAddEndPuncttrue
\mciteSetBstMidEndSepPunct{\mcitedefaultmidpunct}
{\mcitedefaultendpunct}{\mcitedefaultseppunct}\relax
\EndOfBibitem
\bibitem[Zhao and Cieplak(2017)]{yani}
Y.~Zhao and M.~Cieplak, \emph{Phys. Chem. Chem. Phys.}, 2017\relax
\mciteBstWouldAddEndPuncttrue
\mciteSetBstMidEndSepPunct{\mcitedefaultmidpunct}
{\mcitedefaultendpunct}{\mcitedefaultseppunct}\relax
\EndOfBibitem
\bibitem[Poma \emph{et~al.}(2015)Poma, Chwastyk, and Cieplak]{Poma2015}
A.~Poma, M.~Chwastyk and M.~Cieplak, \emph{J. Phys. Chem. B}, 2015,
  \textbf{119}, 12028--12041\relax
\mciteBstWouldAddEndPuncttrue
\mciteSetBstMidEndSepPunct{\mcitedefaultmidpunct}
{\mcitedefaultendpunct}{\mcitedefaultseppunct}\relax
\EndOfBibitem
\bibitem[Cieplak and Hoang(2003)]{Cieplak2003}
M.~Cieplak and T.~X. Hoang, \emph{Biophys. J.}, 2003, \textbf{84},
  475--488\relax
\mciteBstWouldAddEndPuncttrue
\mciteSetBstMidEndSepPunct{\mcitedefaultmidpunct}
{\mcitedefaultendpunct}{\mcitedefaultseppunct}\relax
\EndOfBibitem
\bibitem[Tsai \emph{et~al.}(1999)Tsai, Taylor, Chothia, and Gerstein]{Tsai1999}
J.~Tsai, R.~Taylor, C.~Chothia and M.~Gerstein, \emph{J. Mol. Biol.}, 1999,
  \textbf{290}, 253--66\relax
\mciteBstWouldAddEndPuncttrue
\mciteSetBstMidEndSepPunct{\mcitedefaultmidpunct}
{\mcitedefaultendpunct}{\mcitedefaultseppunct}\relax
\EndOfBibitem
\bibitem[Sulkowska and Cieplak(2008)]{64models}
J.~I. Sulkowska and M.~Cieplak, \emph{Biophys. J.}, 2008, \textbf{95},
  3174--91\relax
\mciteBstWouldAddEndPuncttrue
\mciteSetBstMidEndSepPunct{\mcitedefaultmidpunct}
{\mcitedefaultendpunct}{\mcitedefaultseppunct}\relax
\EndOfBibitem
\bibitem[G{\=o} and Taketomi(1978)]{go1978respective}
N.~G{\=o} and H.~Taketomi, \emph{Proc. Natl. Acad. Sci. U.S.A.}, 1978,
  \textbf{75}, 559--563\relax
\mciteBstWouldAddEndPuncttrue
\mciteSetBstMidEndSepPunct{\mcitedefaultmidpunct}
{\mcitedefaultendpunct}{\mcitedefaultseppunct}\relax
\EndOfBibitem
\bibitem[G{\=o} and Abe(1981)]{go1981noninteracting}
N.~G{\=o} and H.~Abe, \emph{Biopolymers}, 1981, \textbf{20}, 991--1011\relax
\mciteBstWouldAddEndPuncttrue
\mciteSetBstMidEndSepPunct{\mcitedefaultmidpunct}
{\mcitedefaultendpunct}{\mcitedefaultseppunct}\relax
\EndOfBibitem
\bibitem[Hoang and Cieplak(2000)]{Hoang2000}
T.~X. Hoang and M.~Cieplak, \emph{J. Chem. Phys.}, 2000, \textbf{113},
  8319--8328\relax
\mciteBstWouldAddEndPuncttrue
\mciteSetBstMidEndSepPunct{\mcitedefaultmidpunct}
{\mcitedefaultendpunct}{\mcitedefaultseppunct}\relax
\EndOfBibitem
\bibitem[Su\l{}kowska and Cieplak(2007)]{Sulkowska2007}
J.~I. Su\l{}kowska and M.~Cieplak, \emph{J. Phys. Condens. Matter}, 2007,
  \textbf{19}, year\relax
\mciteBstWouldAddEndPuncttrue
\mciteSetBstMidEndSepPunct{\mcitedefaultmidpunct}
{\mcitedefaultendpunct}{\mcitedefaultseppunct}\relax
\EndOfBibitem
\bibitem[Berendsen \emph{et~al.}(1995)Berendsen, van~der Spoel, and van
  Drunen]{Berendsen1995}
H.~Berendsen, D.~van~der Spoel and R.~van Drunen, \emph{Comput. Phys. Commun.},
  1995, \textbf{91}, 43--56\relax
\mciteBstWouldAddEndPuncttrue
\mciteSetBstMidEndSepPunct{\mcitedefaultmidpunct}
{\mcitedefaultendpunct}{\mcitedefaultseppunct}\relax
\EndOfBibitem
\bibitem[van~der Spoel \emph{et~al.}(2005)van~der Spoel, Lindahl, Hess,
  Groenhof, Mark, and Berendsen]{vanderSpoel2005}
D.~van~der Spoel, E.~Lindahl, B.~Hess, G.~Groenhof, A.~Mark and H.~Berendsen,
  \emph{J. Comput. Chem.}, 2005, \textbf{26}, 1701--1718\relax
\mciteBstWouldAddEndPuncttrue
\mciteSetBstMidEndSepPunct{\mcitedefaultmidpunct}
{\mcitedefaultendpunct}{\mcitedefaultseppunct}\relax
\EndOfBibitem
\bibitem[Hess \emph{et~al.}(2008)Hess, Kutzner, van~der Spoel, and
  Lindahl]{Hess2008}
B.~Hess, C.~Kutzner, D.~van~der Spoel and E.~Lindahl, \emph{J. Chem. Theory
  Comput.}, 2008, \textbf{4}, 435--447\relax
\mciteBstWouldAddEndPuncttrue
\mciteSetBstMidEndSepPunct{\mcitedefaultmidpunct}
{\mcitedefaultendpunct}{\mcitedefaultseppunct}\relax
\EndOfBibitem
\bibitem[Pronk \emph{et~al.}(2013)Pronk, P\'all, Schulz, Larsson, Bjelkmar,
  Apostolov, Shirts, Smith, Kasson, van~der Spoel, Hess, and
  Lindahl]{Pronk2013}
S.~Pronk, S.~P\'all, R.~Schulz, P.~Larsson, P.~Bjelkmar, R.~Apostolov,
  M.~Shirts, J.~Smith, P.~Kasson, D.~van~der Spoel, B.~Hess and E.~Lindahl,
  \emph{Bioinformatics}, 2013, \textbf{29}, 845--854\relax
\mciteBstWouldAddEndPuncttrue
\mciteSetBstMidEndSepPunct{\mcitedefaultmidpunct}
{\mcitedefaultendpunct}{\mcitedefaultseppunct}\relax
\EndOfBibitem
\bibitem[Carrion-Vazquez \emph{et~al.}(1999)Carrion-Vazquez, Oberhauser,
  Fowler, Marszalek, Broedel, Clarke, and Fernandez]{Carrion-Vazquez1999}
M.~Carrion-Vazquez, A.~F. Oberhauser, S.~B. Fowler, P.~E. Marszalek, S.~E.
  Broedel, J.~Clarke and J.~M. Fernandez, \emph{Proc. Natl. Acad. Sci. U.S.A.},
  1999, \textbf{96}, 3694--3699\relax
\mciteBstWouldAddEndPuncttrue
\mciteSetBstMidEndSepPunct{\mcitedefaultmidpunct}
{\mcitedefaultendpunct}{\mcitedefaultseppunct}\relax
\EndOfBibitem
\bibitem[Marszalek \emph{et~al.}(1999)Marszalek, Lu, Li, Carrion-Vazquez,
  Oberhauser, Schulten, and Fernandez]{marszalek1999mechanical}
P.~E. Marszalek, H.~Lu, H.~Li, M.~Carrion-Vazquez, A.~F. Oberhauser,
  K.~Schulten and J.~M. Fernandez, \emph{Nature}, 1999, \textbf{402},
  100--103\relax
\mciteBstWouldAddEndPuncttrue
\mciteSetBstMidEndSepPunct{\mcitedefaultmidpunct}
{\mcitedefaultendpunct}{\mcitedefaultseppunct}\relax
\EndOfBibitem
\bibitem[Bahar and Rader(2005)]{Bahar2005}
I.~Bahar and A.~J. Rader, \emph{Curr. Opin. Struct. Biol.}, 2005, \textbf{15},
  586--592\relax
\mciteBstWouldAddEndPuncttrue
\mciteSetBstMidEndSepPunct{\mcitedefaultmidpunct}
{\mcitedefaultendpunct}{\mcitedefaultseppunct}\relax
\EndOfBibitem
\bibitem[Fuglebakk \emph{et~al.}(2013)Fuglebakk, Reuter, and K]{rmsfAA}
E.~Fuglebakk, N.~Reuter and H.~K, \emph{J. Chem. Theory Comput.}, 2013,
  \textbf{9}, 5618--5628\relax
\mciteBstWouldAddEndPuncttrue
\mciteSetBstMidEndSepPunct{\mcitedefaultmidpunct}
{\mcitedefaultendpunct}{\mcitedefaultseppunct}\relax
\EndOfBibitem
\bibitem[Flory(1976)]{Flory1976}
P.~Flory, \emph{Proc. R. Soc. Lond.}, 1976, \textbf{351}, 351--380\relax
\mciteBstWouldAddEndPuncttrue
\mciteSetBstMidEndSepPunct{\mcitedefaultmidpunct}
{\mcitedefaultendpunct}{\mcitedefaultseppunct}\relax
\EndOfBibitem
\bibitem[Valbuena \emph{et~al.}(2009)Valbuena, Oroz, Herv{\'a}s, Vera,
  Rodr{\'\i}guez, Men{\'e}ndez, Sulkowska, Cieplak, and
  Carri{\'o}n-V{\'a}zquez]{Valbuena2009}
A.~Valbuena, J.~Oroz, R.~Herv{\'a}s, A.~M. Vera, D.~Rodr{\'\i}guez,
  M.~Men{\'e}ndez, J.~I. Sulkowska, M.~Cieplak and M.~Carri{\'o}n-V{\'a}zquez,
  \emph{Proc. Natl. Acad. Sci. U.S.A.}, 2009, \textbf{106}, 13791--13796\relax
\mciteBstWouldAddEndPuncttrue
\mciteSetBstMidEndSepPunct{\mcitedefaultmidpunct}
{\mcitedefaultendpunct}{\mcitedefaultseppunct}\relax
\EndOfBibitem
\bibitem[Wlodawer \emph{et~al.}(1986)Wlodawer, Borkakoti, Moss, and
  Howlin]{wlodawer1986comparison}
A.~Wlodawer, N.~Borkakoti, D.~Moss and B.~Howlin, \emph{Acta Crystallogr. B},
  1986, \textbf{42}, 379--387\relax
\mciteBstWouldAddEndPuncttrue
\mciteSetBstMidEndSepPunct{\mcitedefaultmidpunct}
{\mcitedefaultendpunct}{\mcitedefaultseppunct}\relax
\EndOfBibitem
\bibitem[Wlodawer \emph{et~al.}(1984)Wlodawer, Walter, Huber, and
  Sj{\"o}lin]{wlodawer1984structure}
A.~Wlodawer, J.~Walter, R.~Huber and L.~Sj{\"o}lin, \emph{J. Mol. Biol.}, 1984,
  \textbf{180}, 301--329\relax
\mciteBstWouldAddEndPuncttrue
\mciteSetBstMidEndSepPunct{\mcitedefaultmidpunct}
{\mcitedefaultendpunct}{\mcitedefaultseppunct}\relax
\EndOfBibitem
\bibitem[Kabsch \emph{et~al.}(1990)Kabsch, Mannherz, Suck, Pai, and
  Holmes]{kabsch1990atomic}
W.~Kabsch, H.~G. Mannherz, D.~Suck, E.~F. Pai and K.~C. Holmes, \emph{Nature},
  1990, \textbf{347}, 37\relax
\mciteBstWouldAddEndPuncttrue
\mciteSetBstMidEndSepPunct{\mcitedefaultmidpunct}
{\mcitedefaultendpunct}{\mcitedefaultseppunct}\relax
\EndOfBibitem
\bibitem[Cieplak \emph{et~al.}(2002)Cieplak, Hoang, and Robbins]{Cieplak2002}
M.~Cieplak, T.~X. Hoang and M.~O. Robbins, \emph{Proteins: Struct., Funct.,
  Genet.}, 2002, \textbf{49}, 114--124\relax
\mciteBstWouldAddEndPuncttrue
\mciteSetBstMidEndSepPunct{\mcitedefaultmidpunct}
{\mcitedefaultendpunct}{\mcitedefaultseppunct}\relax
\EndOfBibitem
\bibitem[Sotomayor and Schulten(2007)]{sotomayor2007single}
M.~Sotomayor and K.~Schulten, \emph{Science}, 2007, \textbf{316},
  1144--1148\relax
\mciteBstWouldAddEndPuncttrue
\mciteSetBstMidEndSepPunct{\mcitedefaultmidpunct}
{\mcitedefaultendpunct}{\mcitedefaultseppunct}\relax
\EndOfBibitem
\bibitem[Marszalek \emph{et~al.}(1999)Marszalek, Lu, Li, Carrion-Vazquez,
  Oberhauser, Schulten, and Fernandez]{Marszalek1999}
P.~E. Marszalek, H.~Lu, H.~Li, M.~Carrion-Vazquez, A.~F. Oberhauser,
  K.~Schulten and J.~M. Fernandez, \emph{Nature}, 1999, \textbf{402},
  100--103\relax
\mciteBstWouldAddEndPuncttrue
\mciteSetBstMidEndSepPunct{\mcitedefaultmidpunct}
{\mcitedefaultendpunct}{\mcitedefaultseppunct}\relax
\EndOfBibitem
\bibitem[Wojciechowski \emph{et~al.}(2014)Wojciechowski, Szymczak,
  Carri{\'o}n-V{\'a}zquez, and Cieplak]{wojciechowski2014protein}
M.~Wojciechowski, P.~Szymczak, M.~Carri{\'o}n-V{\'a}zquez and M.~Cieplak,
  \emph{Biophys. J.}, 2014, \textbf{107}, 1661--1668\relax
\mciteBstWouldAddEndPuncttrue
\mciteSetBstMidEndSepPunct{\mcitedefaultmidpunct}
{\mcitedefaultendpunct}{\mcitedefaultseppunct}\relax
\EndOfBibitem
\bibitem[Valbuena \emph{et~al.}(2009)Valbuena, Oroz, Herv{\'a}s, Vera,
  Rodr{\'\i}guez, Men{\'e}ndez, Sulkowska, Cieplak, and
  Carri{\'o}n-V{\'a}zquez]{valbuena2009remarkable}
A.~Valbuena, J.~Oroz, R.~Herv{\'a}s, A.~M. Vera, D.~Rodr{\'\i}guez,
  M.~Men{\'e}ndez, J.~I. Sulkowska, M.~Cieplak and M.~Carri{\'o}n-V{\'a}zquez,
  \emph{Proc. Natl. Acad. Sci. U.S.A.}, 2009, \textbf{106}, 13791--13796\relax
\mciteBstWouldAddEndPuncttrue
\mciteSetBstMidEndSepPunct{\mcitedefaultmidpunct}
{\mcitedefaultendpunct}{\mcitedefaultseppunct}\relax
\EndOfBibitem
\bibitem[Fowler \emph{et~al.}(2002)Fowler, Best, Herrera, Rutherford, Steward,
  Paci, Karplus, and Clarke]{fowler2002mechanical}
S.~B. Fowler, R.~B. Best, J.~L.~T. Herrera, T.~J. Rutherford, A.~Steward,
  E.~Paci, M.~Karplus and J.~Clarke, \emph{J. Mol. Biol.}, 2002, \textbf{322},
  841--849\relax
\mciteBstWouldAddEndPuncttrue
\mciteSetBstMidEndSepPunct{\mcitedefaultmidpunct}
{\mcitedefaultendpunct}{\mcitedefaultseppunct}\relax
\EndOfBibitem
\bibitem[Best \emph{et~al.}(2003)Best, Fowler, Herrera, Steward, Paci, and
  Clarke]{best2003mechanical}
R.~B. Best, S.~B. Fowler, J.~L.~T. Herrera, A.~Steward, E.~Paci and J.~Clarke,
  \emph{J. Mol. Biol.}, 2003, \textbf{330}, 867--877\relax
\mciteBstWouldAddEndPuncttrue
\mciteSetBstMidEndSepPunct{\mcitedefaultmidpunct}
{\mcitedefaultendpunct}{\mcitedefaultseppunct}\relax
\EndOfBibitem
\bibitem[Schwaiger \emph{et~al.}(2005)Schwaiger, Schleicher, Noegel, and
  Rief]{schwaiger2005folding}
I.~Schwaiger, M.~Schleicher, A.~A. Noegel and M.~Rief, \emph{EMBO reports},
  2005, \textbf{6}, 46--51\relax
\mciteBstWouldAddEndPuncttrue
\mciteSetBstMidEndSepPunct{\mcitedefaultmidpunct}
{\mcitedefaultendpunct}{\mcitedefaultseppunct}\relax
\EndOfBibitem
\bibitem[Schwaiger \emph{et~al.}(2004)Schwaiger, Kardinal, Schleicher, Noegel,
  and Rief]{schwaiger2004mechanical}
I.~Schwaiger, A.~Kardinal, M.~Schleicher, A.~A. Noegel and M.~Rief, \emph{Nat.
  Struct. Mol. Biol.}, 2004, \textbf{11}, 81\relax
\mciteBstWouldAddEndPuncttrue
\mciteSetBstMidEndSepPunct{\mcitedefaultmidpunct}
{\mcitedefaultendpunct}{\mcitedefaultseppunct}\relax
\EndOfBibitem
\bibitem[Frantz(2009)]{frantz2009g3data}
J.~Frantz, \emph{URL: http://www.frantz.fi/software/g3data.php/ Version 1},
  2009\relax
\mciteBstWouldAddEndPuncttrue
\mciteSetBstMidEndSepPunct{\mcitedefaultmidpunct}
{\mcitedefaultendpunct}{\mcitedefaultseppunct}\relax
\EndOfBibitem
\bibitem[Sikora \emph{et~al.}(2009)Sikora, Su{\l}kowska, and
  Cieplak]{sikora2009mechanical}
M.~Sikora, J.~I. Su{\l}kowska and M.~Cieplak, \emph{PLOS Comput. Biol.}, 2009,
  \textbf{5}, e1000547\relax
\mciteBstWouldAddEndPuncttrue
\mciteSetBstMidEndSepPunct{\mcitedefaultmidpunct}
{\mcitedefaultendpunct}{\mcitedefaultseppunct}\relax
\EndOfBibitem
\bibitem[Kouza \emph{et~al.}(2009)Kouza, Hu, Zung, and Li]{kouza2009protein}
M.~Kouza, C.-K. Hu, H.~Zung and M.~S. Li, \emph{J. Chem. Phys.}, 2009,
  \textbf{131}, 12B608\relax
\mciteBstWouldAddEndPuncttrue
\mciteSetBstMidEndSepPunct{\mcitedefaultmidpunct}
{\mcitedefaultendpunct}{\mcitedefaultseppunct}\relax
\EndOfBibitem
\bibitem[van Nuland \emph{et~al.}(1994)van Nuland, Hangyi, van Schaik,
  Berendsen, van Gunsteren, Scheek, and Robillard]{van1994high}
N.~A. van Nuland, I.~W. Hangyi, R.~C. van Schaik, H.~J. Berendsen, W.~F. van
  Gunsteren, R.~M. Scheek and G.~T. Robillard, \emph{J. Mol. Biol.}, 1994,
  \textbf{237}, 544--559\relax
\mciteBstWouldAddEndPuncttrue
\mciteSetBstMidEndSepPunct{\mcitedefaultmidpunct}
{\mcitedefaultendpunct}{\mcitedefaultseppunct}\relax
\EndOfBibitem
\bibitem[Gronenborn \emph{et~al.}(1991)Gronenborn, Filpula, Essig, Achari,
  Whitlow, Wingfield, and Clore]{gronenborn1991novel}
A.~M. Gronenborn, D.~R. Filpula, N.~Z. Essig, A.~Achari, M.~Whitlow, P.~T.
  Wingfield and G.~M. Clore, \emph{Science}, 1991, \textbf{253}, 657--661\relax
\mciteBstWouldAddEndPuncttrue
\mciteSetBstMidEndSepPunct{\mcitedefaultmidpunct}
{\mcitedefaultendpunct}{\mcitedefaultseppunct}\relax
\EndOfBibitem
\bibitem[Munoz \emph{et~al.}(1997)Munoz, Thompson, Hofrichter, and
  Eaton]{Munoz_Nature1997}
V.~Munoz, P.~A. Thompson, J.~Hofrichter and W.~A. Eaton, \emph{Nature}, 1997,
  \textbf{390}, 196--199\relax
\mciteBstWouldAddEndPuncttrue
\mciteSetBstMidEndSepPunct{\mcitedefaultmidpunct}
{\mcitedefaultendpunct}{\mcitedefaultseppunct}\relax
\EndOfBibitem
\bibitem[Kubelka \emph{et~al.}(2004)Kubelka, Hofrichter, and
  Eaton]{kubelka2004protein}
J.~Kubelka, J.~Hofrichter and W.~A. Eaton, \emph{Curr. Opin. Struct. Biol.},
  2004, \textbf{14}, 76--88\relax
\mciteBstWouldAddEndPuncttrue
\mciteSetBstMidEndSepPunct{\mcitedefaultmidpunct}
{\mcitedefaultendpunct}{\mcitedefaultseppunct}\relax
\EndOfBibitem
\bibitem[Nguyen \emph{et~al.}(2005)Nguyen, Stock, Mittag, Hu, and
  Li]{Phuong_Proteins2005}
P.~H. Nguyen, G.~Stock, E.~Mittag, C.~K. Hu and M.~S. Li, \emph{Proteins},
  2005, \textbf{61}, 795--808\relax
\mciteBstWouldAddEndPuncttrue
\mciteSetBstMidEndSepPunct{\mcitedefaultmidpunct}
{\mcitedefaultendpunct}{\mcitedefaultseppunct}\relax
\EndOfBibitem
\bibitem[Zagrovic \emph{et~al.}(2001)Zagrovic, Sorin, and
  Pande]{Zagrovic_JMB2001}
B.~Zagrovic, E.~J. Sorin and V.~Pande, \emph{J. Mol. Biol.}, 2001,
  \textbf{313}, 151--169\relax
\mciteBstWouldAddEndPuncttrue
\mciteSetBstMidEndSepPunct{\mcitedefaultmidpunct}
{\mcitedefaultendpunct}{\mcitedefaultseppunct}\relax
\EndOfBibitem
\bibitem[Garcia and Sanbonmatsu(2001)]{Garcia_Proteins2001}
A.~E. Garcia and K.~Y. Sanbonmatsu, \emph{Proteins}, 2001, \textbf{42},
  345--354\relax
\mciteBstWouldAddEndPuncttrue
\mciteSetBstMidEndSepPunct{\mcitedefaultmidpunct}
{\mcitedefaultendpunct}{\mcitedefaultseppunct}\relax
\EndOfBibitem
\end{mcitethebibliography}
\bibliographystyle{rsc} 

\end{document}